\documentclass[aps,10pt,pre,floatfix,twocolumn,superscriptaddress]{revtex4-1}

\usepackage{color}
\usepackage{ulem}
\usepackage{graphicx}
\usepackage{subfigure,amsmath, amsthm, amssymb}
\usepackage[colorlinks,linkcolor=blue,citecolor=blue]{hyperref}
\usepackage{csquotes}
\graphicspath{{./Figures/}}

\newcommand{\be}{\begin{equation}}
\newcommand{\ee}{\end{equation}}
\newcommand{\bea}{\begin{eqnarray}}
\newcommand{\eea}{\end{eqnarray}}

\begin{document}

\title{Mpemba effect in driven granular Maxwell gases}

\author{Apurba Biswas}
\email{apurbab@imsc.res.in}
\affiliation{The Institute of Mathematical Sciences, C.I.T. Campus, Taramani, Chennai 600113, India}
\affiliation{Homi Bhabha National Institute, Training School Complex, Anushakti Nagar, Mumbai 400094, India}
\author{V . V. Prasad}
\email{prasadcalicut@gmail.com}
\affiliation{Department of Physics of Complex Systems, Weizmann Institute of Science, 76100 Rehovot, Israel}
\affiliation{Government Arts and Science College, Nadapuram, Kozhikode 673506, India}
\author{O. Raz}
\email{oren.raz@weizmann.ac.il}
\affiliation{Department of Physics of Complex Systems, Weizmann Institute of Science, 76100 Rehovot, Israel}
\author{R. Rajesh} 
\email{rrajesh@imsc.res.in}
\affiliation{The Institute of Mathematical Sciences, C.I.T. Campus, Taramani, Chennai 600113, India}
\affiliation{Homi Bhabha National Institute, Training School Complex, Anushakti Nagar, Mumbai 400094, India}

\begin{abstract}
 A Mpemba effect refers to the counterintuitive result that, when quenched to a low temperature, a system at higher temperature  may equilibrate faster than one at  intermediate temperatures. This effect has recently been demonstrated in driven granular gases, both for
smooth as well as rough hard-sphere systems based on a perturbative analysis. In this paper, we consider the inelastic driven Maxwell gas, a simplified model for a granular gas, where the rate of collision is assumed to be independent of the relative velocity.  Through an exact analysis, we determine the conditions under which the Mpemba effect is present in this model. For mono-dispersed gases, we show that 
the Mpemba effect is present only when the initial states are allowed to be non-stationary, while for bi-dispersed gases, it is present for some 
steady state initial states.  We also demonstrate the  existence of the strong  Mpemba effect for bi-dispersed Maxwell gas wherein the system at higher temperature relaxes to a final steady state at an exponentially faster rate leading to smaller equilibration time.
\end{abstract}

%\keywords{}

\maketitle
\section{Introduction}

A classic lore associated with everyday experience is that under certain conditions, hot water freezes faster than cold water. 
Even though this intriguing effect  has been known since Aristotelian times~\cite{aristotle,1952meteorologica},
it has been named the Mpemba effect after E. B.  Mpemba, 
who was the first to study it systematically~\cite{Mpemba_1969}.
This out of equilibrium relaxation behavior is in stark contrast to quasi-static relaxations, where systems with different initial conditions 
evolve in time through equilibrium states and therefore their temperatures never cross.

Several mechanisms were suggested to explain the Mpemba effect in water. These include conceptually simple mechanisms, as supercooling~\cite{david-super-cooling-1995}, convection~\cite{vynnycky-convection:2015} and evaporation~\cite{Mirabedin-evporation-2017}, as well as more sophisticated explanations as the anomalous relaxation of the hydrogen bond~\cite{zhang-hydrbond1-2014,tao-hydrogen-2017}. Others have argued that the effect does not exist in water~\cite{NoMpemba}. Regardless of the true status of the effect in water, a similar  effect was experimentally observed in a wide range of physical systems ranging from magnetic alloys~\cite{chaddah2010overtaking} and clathrate hydrates~\cite{paper:hydrates} to polylactides~\cite{Polylactide}.

In addition to experiments, various numerical tools and model systems were  used  to shed light on such anomalous relaxations. To better understand the effect in water, detailed molecular dynamic simulations were performed  \cite{gijon2019paths,zhang-hydrbond1-2014,Molecular_Dynamics_jin2015mechanisms}. In these simulations, the system is made of up to several thousand molecules, with pairwise interactions that model the interactions between the water molecules. The initial configuration of the system is commonly sampled from the Boltzmann distribution of the hot or warm temperature, and the dynamic follows a standard molecular dynamic protocol that corresponds to the cold temperature. A different numerical approach was applied on spin systems, for both ordered~\cite{PhysRevLett.124.060602} as well as glassy~\cite{SpinGlassMpemba} models. Using a Monte Carlo simulation, the value of some order parameters were tracked during the relaxation process of systems that were sampled from the equilibrium distribution of a hot and a warm temperature, and evolved under Markovian dynamics that corresponds to a cold temperature. If, during the relaxation process the corresponding order parameters of the hot and cold systems intersect, then the Mpemba effect exists in the system. 

On the analytic front, two different approaches were used to address the Mpemba effect so far. In Refs.~\cite{Lu-raz:2017,Klich-2019,klich2018solution}, the Mpemba effect was defined and evaluated through the distance between probability distributions during the relaxation process. To this end, the probability distribution describing the system is initiated at the equilibrium (Boltzmann) distribution of the hot or warm temperature. The system is then quenched into an environment with a cold temperature, and thus the probability distribution of the system evolves in time towards the equilibrium of the cold temperature. By tracking the distance, in probability space, between the time dependent probability and the final equilibrium,  the Mpemba effect can be identified, and the exact conditions for its existence were derived. Moreover, this framework naturally suggest a rich class of related phenomena, including the inverse Mpemba effect~\cite{Lu-raz:2017} where a cold system heats up faster than a warmer system; the strong Mpemba effect~\cite{Klich-2019} where a specific initial temperature results in a jump in the relaxation rate; and non-monotonic optimal heating protocols~\cite{PhysRevLett.124.060602} where the optimal heating protocol has a pre-cooling stage.

A different theoretical framework was used in the context of driven granular gases~\cite{Lasanta-mpemba-1-2017,Torrente-rough-2019,mompo2020memory}. A granular gas is a dilute composition of  particles that move ballistically and interact through momentum conserving binary inelastic collisions. These dissipative systems approach a steady state when externally driven to compensate for the kinetic 
energy  which is lost in the inter-particle collisions. For the driven granular system with smooth mono-dispersed particles 
which interact via binary collisions with a rate proportional to the magnitude of the relative velocities of the colliding particles, 
it was shown that the total energy of a system with a higher initial energy attains the final low energy state before a similar system
with an intermediate initial energy~\cite{Lasanta-mpemba-1-2017}. The effect was achieved by independently varying the deviation of the velocity distribution from a Gaussian, characterized by the coefficient of the second Sonine polynomial $a_2$.  In this approach, the system was initiated in a non-stationary distribution, and relaxed towards the corresponding steady state (which is not an equilibrium distribution) associated with the parameters of the system. The Mpemba effect then exists in the system if a non-equilibrium system which is further away from equilibrium, namely has more energy in its initial state, equilibrates faster than an initial condition which is closer to the equilibrium state.  The Mpemba effect was also demonstrated for a rough granular gas, where a much larger range of initial energies result in anomalous relaxations~\cite{Torrente-rough-2019} as well as  for a gas of viscoelastic particles~\cite{mompo2020memory}. In both the rough and smooth granular gas,  the velocity distribution at all times was approximated by a  Gaussian or Gaussian and first order corrections respectively, making the calculations perturbative in nature.

Do the results in the various frameworks for different systems correspond to the same effect? This is a key question, for which the answer is yet unknown. In this manuscript, we partially address a specific difference that plays an important role in both the numerical and analytical results developed so far -- the initial condition of the system. In the Markovian framework, the initial condition of the hot and warm systems is an equilibrium distribution corresponding to the initial temperature~\cite{Lu-raz:2017,Klich-2019,klich2018solution}. In contrast, in the granular gas approach the initial distributions are not a steady state of the system for any (effective) temperature, but are rather transient distributions with different amount of total energy. These distributions relax towards the steady-state distribution by energy exchange between the particles as well as with some bath~\cite{Lasanta-mpemba-1-2017,Torrente-rough-2019}. A similar difference in the initial conditions exists between the various molecular dynamic simulations calculated for water molecules: in \cite{Molecular_Dynamics_jin2015mechanisms} the system was initially sampled from a hot or warm temperature and then quenched to a cold environment, whereas in~\cite{gijon2019paths} the initial condition is not the equilibrium distribution of any temperature, but rather an altered distribution of the final temperature.

To address this specific difference between the two approaches, we consider the inelastic Maxwell model for granular gas in which the collision rates are assumed to be independent of the relative velocity~\cite{Ben-naim:00,Bobylev:00}. The model thus presents a simpler system of granular gas keeping the essential physics intact, while allowing for exact calculation. We investigate the existence of the Mpemba effect for both mono-dispersed and bi-dispersed systems. In Maxwell gases, the equations for the relevant two-point correlations is known to form a closed set of equations~\cite{prasad2014high,Prasad:14,Biswas_2020}. By analyzing these equations in detail, we determine the parameter regime in which  the Mpemba effect can be seen in these systems. In particular, we show that while transient initial conditions are required for the effect to be present in mono-dispersed gas, the bi-dispersed system shows   the Mpemba effect for  steady state initial conditions. This allows us to use the Markovian approach for the Mpemba effect in driven granular gas, and therefore identify the existence of the strong effect in this system, where for a specific initial steady state the relaxation rate is smaller than from any other initial steady state.

The remainder of the  paper is organized as follows. We define the model in Sec.~\ref{model}. In Sec.~\ref{two point corr}, we show that the time evolution of the two-point velocity-velocity correlation functions do not involve higher order correlations and form a closed set of equations.  The time evolution of the two point correlations thus have exact solutions. In Sec.~\ref{sec: mono-dispersed gas}, we define  the Mpemba effect and demonstrate its existence  in driven mono-dispersed systems. Section~\ref{bidispersed} contains the detailed analysis for the existence of  the Mpemba effect  in bi-dispersed granular systems. We also demonstrate the existence of  the inverse Mpemba effect and a stronger version of  the Mpemba effect in similar systems. Section~\ref{conclusion} contains the 
summary of results and a discussion of their implications.

\section{Model \label{model}}

In this paper, we analyze  both mono-dispersed as well as bi-dispersed driven inelastic Maxwell gases.  We first define the bi-dispersed gas, and then indicate the limits when it reduces to a  mono-dispersed gas. Consider $N_A$ particles of type  $A$, each of mass $m_A$, and $N_B$ particles of type $B$, each of mass $m_B$. Let $N_A+N_B=N$.   Each particle has a scalar velocity  $v_{i,k}$, where  $i=1,\ldots, N$ and $k \in\{A,B\}$. These velocities  evolve in time through binary collisions and external driving.  A pair of particles of type $k$ and $l$, 
where $k,l\in\{A,B\}$, collide with rate ${\lambda_{kl}}/N$. The factor $1/N$ in the collision rates ensures that the total rate of collisions between $N_k[N_k-1]/2$ pairs of similar type of particles and that between $N_AN_B$ pairs of different type of particles  are proportional 
to the system size $N$. During a collision, momentum is conserved, but energy is dissipated. Let $v_{i,k}$ and $v_{j,l}$ denote the pre-collision velocities and $v'_{i,k}$, $v'_{j,l}$ 
denote the post-collision velocities. Then
\begin{align}
    v'_{i,k} &= v_{i,k} - (1+r_{kl})\frac{m_l}{m_k + m_l}(v_{i,k} - v_{j,l}), \nonumber \\  
    v'_{j,l} &= v_{j,l} + (1+r_{kl})\frac{m_k}{m_k + m_l}(v_{i,k} - v_{j,l}), 
        \label{collision}
\end{align}
where $k, l=A, B$,  $r_{kl} \in [0,1]$ is the coefficient of restitution for the collision, and $m_k$ and $m_l$ are the masses. There are three coefficients of restitution: $r_{AA}$, $r_{BB}$, and $r_{AB}$ depending on whether the pair of colliding particles are of type \textit{AA}, \textit{BB}, or \textit{AB}. It is convenient to define
\begin{equation}
\alpha_{kl}= \frac{1+r_{kl}}{2}, ~~k, l = A, B,
\label{eq:alpha}
\end{equation}
where $1/2\leq \alpha_{kl}\leq 1$.

In addition to collisions, the system evolves through external driving. We implement a driving scheme that drives the system to a steady state, and has been used in earlier studies~\cite{prasad2014high,Prasad:14,prasad2017velocity}.  Each particle is driven at a rate $\lambda_d$. During such an event, the velocity of the driven particle is modified according to
\begin{equation}
v'_{i,k}=-r_w v_{i,k} + \eta_{k}, \quad -1<r_w \leq 1,~~~ k=A, B,  \label{driving}
\end{equation}
where $r_w \in (-1,1]$ is a parameter and $\eta_k$ is noise drawn from a fixed distribution $\phi_k(\eta_k)$.   
There is no compelling reason for  $\phi_k(\eta_k)$ to be Gaussian. However, we restrict ourselves to distributions with zero mean and finite second moment $\sigma^2_k$ given by
\be
\sigma^2_k= \int_{-\infty}^{\infty} d\eta~ \eta^2 \phi_k(\eta), ~~~ k=A, B.
\label{eq:noisevariance}
\ee
The physical motivations for the form of driving may be found in Refs.~\cite{prasad2019asymptotic,Prasad_2019}.
 Without loss of generality, in all the plots, we set the driving rate $\lambda_d=1$, so that time is measured in units of $\lambda_d^{-1}$.

In the model, the spatial degrees of freedom have been neglected. This corresponds to the well-mixed limit where the spatial correlations between particles are ignored. In addition, we have assumed that the collision rates are independent of the relative velocity of the colliding particles. This corresponds to the so-called Maxwell limit.

Let $P_k(v,t)$, where $k=A,B$,  denote the probability that  a randomly chosen particle of type $k$ has velocity $v$ at  time $t$. Its time evolution is given by:
\begin{widetext}
 \begin{align}
           & \frac{d}{dt}P_k(v,t) = \frac{\lambda_{kk}(N_k-1)}{N} \int \int dv_1 dv_2 P_{k}(v_1, t) P_{k}(v_2, t) \delta[ ( 1-\alpha_{kk})v_1 +  \alpha_{kk} v_2-v]\nonumber \\
           & + \frac{\lambda_{k\bar{k}}N_{\bar{k}}}{N} \int \int dv_1 dv_2 P_{k}(v_1, t) P_{\bar{k}}(v_2, t) \delta[ ( 1-X_{\bar{k}})v_1 +  X_{\bar{k}}v_2-v] -\frac{\lambda_{kk}(N_k-1)}{N}  P_{k}(v, t)  \nonumber \\
           & -\frac{\lambda_{k\bar{k}} N_{\bar{k}}}{N}  P_{k}(v, t) +\lambda_d\left[ - P_k(v, t) + \int \int d\eta_k dv_1 \phi_k(\eta_k) P_k(v_1, t) \delta[ -r_wv_1 + \eta_k - v]\right], 
           \label{pa_single}
\end{align}
\end{widetext}
 where 
\begin{equation}
\bar{k}=
\begin{cases}
B,\quad   \text{if}\quad k=A,   \\
A,\quad  \text{if}\quad k=B,     \label{eq:particle label}
\end{cases}
\end{equation}
and
\begin{align}
X_{k} =\alpha_{AB} \mu_k~\text{where}~\mu_k=\frac{2m_{k}}{m_A+m_B}, ~ k = A, B,
\label{eq: redefine x}
\end{align}
with $\mu_k\in(0,2)$ and $\alpha_{AB}$ is defined in Eq.~(\ref{eq:alpha}).

The mono-dispersed Maxwell gas is obtained by taking the limit $N_A=N$ with $(N_A-1)/N \rightarrow 1$, $N_B=0$, $r_{AA}=r$ and setting all other coefficients of restitution to zero. The rate of inter-particle collisions is denoted by $\lambda$ and the rate of driving for the particles by $\lambda_d$. If $P(v,t)$ denote the probability that a randomly chosen particle has velocity $v$ at time $t$ then its time evolution, for a mono-dispersed gas is given by:
\begin{align}
&\frac{d}{dt}P(v,t)=-\lambda P(v,t)- \lambda_d P(v, t) \nonumber \\
&+ \lambda \int \int dv_1 dv_2 P(v_1, t) P(v_2, t) \delta[ ( 1-\alpha)v_1 +  \alpha v_2-v]\nonumber \\
&+ \lambda_d \int \int d\eta dv_1 \phi(\eta) P(v_1, t) \delta[ -r_wv_1 + \eta - v],
\label{eq:monodispersed}
\end{align}
where 
\begin{align}
\alpha=\frac{1+r}{2}.
\end{align}

\section{Calculation of the two point correlations \label{two point corr}}

In this section, we define the relevant  two point correlation functions for both mono-dispersed and bi-dispersed gases. The evolution equations for these correlation functions were derived in Refs.~\cite{prasad2014high,Biswas_2020}. We summarize these derivations and then develop a solution that will be useful for demonstrating  the Mpemba effect. Sections~\ref{mono-dispersed} and \ref{bi-dispersed} contain the derivation for mono-dispersed and bi-dispersed gases respectively.

\subsection{Mono-dispersed Maxwell gas \label{mono-dispersed}}

We first discuss the case of mono-dispersed Maxwell gas. Consider the following two point correlation functions:
\begin{align}
\begin{split}
&E(t)=\frac{1}{N}\sum^{N}_{i=1} \langle v^2_i(t)\rangle, \\
&C(t)=\frac{1}{N(N-1)}\sum^{N}_{i=1} \sum^N_{j=1,j\not =i} \langle v_{i}(t) v_{j}(t) \rangle,
\label{2 point correlations monodispersed}
\end{split}
\end{align}
where $E(t)$ is the mean kinetic energy of a particle, and $C(t)$ is the equal time velocity-velocity correlation between a pair of particles. In the steady state, the  inter-particle two point correlation function is known to be zero~\cite{Prasad:14}. However, for the purpose of demonstrating   the Mpemba effect, we consider non-zero correlations, which  correspond to non-stationary states. The time evolution of these correlation functions can be obtained in a straightforward manner from Eq.~(\ref{eq:monodispersed}) and 
can be compactly represented in matrix form as~\cite{prasad2014high}
\begin{align}
\frac{d\boldsymbol{\Sigma}(t)}{dt} =-\boldsymbol{R}  \boldsymbol{\Sigma}(t) + \boldsymbol{D},
\label{two-point-correlation-matrix-evolution monodispersed}
\end{align}
where $\boldsymbol{\Sigma}(t)=[E(t),C(t)]^T$,  $\boldsymbol{D}=[\lambda_d \sigma^2, 0]^T$,
 and $\boldsymbol{R}$ is given by
{\small
\begin{equation}
\boldsymbol{R}\!=\!\left[\begin{array}{cc} 
\!\lambda_c(1-r^2) \!+ \! \lambda_d(1-r_w^2)&-\lambda_c(1-r^2)\\
-\frac{\lambda_c(1-r^2)}{N-1}&\frac{\lambda_c(1-r^2)}{N-1}\!+\!\lambda_d(1+r_w)
\end{array}\right].
\label{eq:Rmono}
\end{equation}}
Note that while $R_{11}$ and $R_{22}$ are positive, $R_{12}$ and $R_{21}$ are negative.

Equation~(\ref{two-point-correlation-matrix-evolution monodispersed}) for the correlation functions  forms a closed set of linear equations and does not involve higher order correlation functions. This allows for a complete solution.
Equation~(\ref{two-point-correlation-matrix-evolution monodispersed}) can be solved exactly by linear decomposition using 
the eigenvalues $\lambda_{\pm}$ of $\boldsymbol{R}$:
{\small
\begin{align}
\lambda_\pm=\frac{R_{11}+R_{22} \pm\sqrt{(R_{11}-R_{22})^2+4 R_{21}R_{12}}}{2}.
\end{align}}
Here, the eigenvalues $\lambda_{\pm}>0$ with $\lambda_+>\lambda_-$.
The solution for  $E(t)$ and $C(t)$ can then be obtained as:
\begin{align}
\begin{split}
E(t)&= K_+ e^{-\lambda_+t}+K_- e^{-\lambda_-t }+K_0,\\
C(t)&= L_+ e^{-\lambda_+t}+L_- e^{-\lambda_-t }+L_0,
\end{split}
\label{time evolution exact monodispersed}
\end{align}
where the coefficients $K_+, K_-,K_0, L_+, L_-$ and $L_0$ are given in Eq.~(\ref{appendix coeff monodispersed}) of Appendix~\ref{appendix two point correlation}.

\subsection{Bi-dispersed Maxwell gas \label{bi-dispersed}}

For the case of bi-dispersed Maxwell gas, we can define  two point correlation functions, $E_A$ and $E_B$  for the mean kinetic energies of type $A$ and $B$ particles, and three two point velocity-velocity correlation functions $C_{ij}$, where $i,j \in (A,B)$:
\begin{align}
\begin{split}
E_A(t)&=\frac{1}{N_A}\sum^{N_A}_{i=1} \langle v^2_{i,A}(t) \rangle, \\
E_B(t)&=\frac{1}{N_B}\sum^{N_B}_{i=1} \langle v^2_{i,B}(t) \rangle,\\
C_{AB}(t)&=\frac{1}{N_A N_B}\sum^{N_A}_{i=1} \sum^{N_B}_{j=1} \langle v_{i,A}(t) v_{j,B}(t) \rangle, \\
C_{AA}(t)&=\frac{1}{N_A (N_A-1)}\sum^{N_A}_{i=1} \sum_{\substack{j=1 \\ j \neq i}}^{N_A} \langle v_{i,A}(t) v_{j,A}(t) \rangle, \\
C_{BB}(t)&=\frac{1}{N_B (N_B-1)}\sum^{N_B}_{i=1} \sum_{\substack{j=1 \\ j \neq i}}^{N_B} \langle v_{i,B}(t) v_{j,B}(t) \rangle. \label{2 point correlations}
\end{split}
\end{align} 
The time evolution for these correlation function can be obtained from  Eq.~(\ref{pa_single}), as  derived in Ref.~\cite{Biswas_2020} when only one type of particle is driven. We generalize these calculations to the case where both types of particles are driven. The time evolution equations are linear and form a closed set of equations as shown in Appendix~(\ref{appendix a bidispersed}).

In the steady state, in the thermodynamic limit, the inter-particle two point correlation functions $C_{ij}$, where $i,j \in (A,B)$ are zero, as shown in Ref.~\cite{Biswas_2020}. If in the initial state, these correlations are zero, then it remains zero for all times. We will be only considering such initial states. Unlike the mono-dispersed case, we will show that  the Mpemba effect is possible for initial states that are steady states. In that case, we can ignore these correlations, and
write the time evolution of mean kinetic energies of $A$ and $B$ type particles, i.e.,  $E_A$ and $E_B$ respectively  in a compact form (see Appendix~\ref{appendix a bidispersed} for detailed calculations) as
\begin{equation}
\frac{d\boldsymbol{\Sigma}(t)}{dt}=~\boldsymbol{R}\boldsymbol{\Sigma}(t)+\boldsymbol{D},\label{time-ev}
\end{equation}
where
\begin{align}
\boldsymbol{\Sigma}(t)&=\begin{bmatrix}
E_{A}(t), &
E_{B}(t)
\end{bmatrix}^T, \\
\boldsymbol{D}&=\begin{bmatrix}
\lambda_d \sigma^2_A, &
\lambda_d \sigma^2_B
\end{bmatrix}^T,
\end{align}
and  $\boldsymbol{R}$ is a  $2\times2$ matrix, whose entries are given by
\begin{align}
\begin{split}
R_{11}=&\lambda_{AB}\nu_B X^2_B-\lambda_{AA}\alpha_{AA}(1-\alpha_{AA})\nu_A \\ 
& - 2\lambda_{AB}\nu_B X_B-\lambda_d(1-r^2_w),\\ 
R_{12}=&\lambda_{AB}\nu_B X^2_B, \\
R_{21}=&\lambda_{AB}\nu_A X^2_A, \\ 
R_{22}=&\lambda_{AB}\nu_A X^2_A-\lambda_{BB}\alpha_{BB}(1-\alpha_{BB})\nu_B \\
&- 2\lambda_{AB}\nu_A X_A-\lambda_d(1-r^2_w),
\end{split}
\end{align}
 where $\nu_A$ and $\nu_B$ are the fraction of $A$ and $B$ type particles, respectively.  We note that the linearity in the evolution equations [Eqs.~(\ref{two-point-correlation-matrix-evolution monodispersed}) and (\ref{time-ev})] for the energy arises naturally for Maxwell gases (both mono-dispersed and bi-dispersed) when compared to the  granular gas models studied earlier~\cite{Lasanta-mpemba-1-2017,Torrente-rough-2019},
wherein  the non-linear evolution equation limits analytical treatment unless linearized using perturbative methods~\cite{Lasanta-mpemba-1-2017}. 
Similar exact linear evolution equation 
has been analyzed in the case of Markovian Mpemba effect~\cite{Lu-raz:2017} where the vector denoting the probabilities of various states evolves according to an equation similar to Eqs.~(\ref{two-point-correlation-matrix-evolution monodispersed}) and (\ref{time-ev}).

\section{The Mpemba effect in Mono-dispersed Maxwell gas \label{sec: mono-dispersed gas}}

In this section, we derive the conditions for   the Mpemba effect to be present in the mono-dispersed driven Maxwell gas, based on an analysis of Eq.~(\ref{time evolution exact monodispersed}) for the solution of $E(t)$ and $C(t)$.  More precisely, we   define   the Mpemba effect as follows.  Consider two systems with two different granular temperatures, or kinetic energies [the terms ``granular temperature" and ``kinetic energies" are used interchangeably]. We let these systems evolve to a steady state at the same final temperature, that is lower than the initial temperatures. If the hotter system cools faster (the energy-time plots show a crossing), then we say that the system shows   the Mpemba effect.

We now proceed to find out the criteria for   the Mpemba effect to be present in the 
mono-dispersed Maxwell gas. Consider two systems labeled as $P$ and $Q$. Let their initial conditions be denoted by
$(E^P(0),C^P(0))$ and $(E^Q(0),C^Q(0))$ with $E^P(0)>E^Q(0)$. Both systems are then driven to a common steady state. This is achieved when the systems $P$ and $Q$ are driven with the same driving strength ($\sigma$) which is chosen such that the mean kinetic energy of the common steady state is lower than the initial mean kinetic energies of $P$ and $Q$,  while keeping all the other parameters of both the systems constant.

If this system shows   a Mpemba effect, then the trajectories  $E^P(t)$ and $E^Q(t)$ must cross each other, such that  there is a 
time $t=\tau$ at which
\begin{align}
E^P(\tau)=E^Q(\tau).
\end{align}
Substituting into Eq.~(\ref{time evolution exact monodispersed}), we obtain
relation:
\begin{align}
K^P_+ e^{-\lambda_+\tau}+K^P_- e^{-\lambda_-\tau }=K^Q_+ e^{-\lambda_+\tau}+K^Q_- e^{-\lambda_-\tau},
\end{align}
whose solution is
\bea
\tau=\frac{1}{\lambda_+-\lambda_-}\ln\left[\frac{K_+^P-K_+^Q}{K_-^P-K_-^Q}\right],
\eea
which in terms of the initial conditions reduce to
\bea
\tau=\frac{1}{\lambda_+-\lambda_-}\ln\left[\frac{R_{12}\Delta C-(\lambda_--R_{11})\Delta E}{R_{12}\Delta C-(\lambda_+-R_{11})\Delta E}\right],
\label{mpemba monodispersed}
\eea
where 
\bea
\Delta E &=& E^P(0)-E^Q(0),\\
\Delta C &=& C^P(0)-C^Q(0).
\eea

For  the Mpemba effect to be present, we require that $\tau >0$. Since $\lambda_+> \lambda_-$,  the argument of logarithm in Eq.~(\ref{mpemba monodispersed}) should be greater than one. We immediately obtain the criterion 
 \begin{align}
(\lambda_+-R_{11})\Delta E<R_{12}\Delta C.
\label{condition mpemba monodispersed}
\end{align}
Note that $R_{12}<0$, and $\lambda_+>R_{11}$ [see Eq.~(\ref{eq:Rmono})]. Since $\Delta E >0$ by definition, we conclude that  Eq.~(\ref{condition mpemba monodispersed}) can be satisfied  only if $\Delta C<0$, i.e.,  the two point velocity-velocity correlation of 
the hotter initial system  $C^P(0)$ is sufficiently smaller than that of  the cooler counterpart 
$C^Q(0)$.  Note that if the two systems $P$ and $Q$ were initially in a steady state, then in the thermodynamic limit the correlations vanish, i.e.,  $\Delta C=\mathcal{O}(1/N)$, and the inequality in Eq.~(\ref{condition mpemba monodispersed}) cannot be satisfied. Thus, for   the Mpemba effect to be present in the mono-dispersed gas, the initial condition of the cooler component cannot be a steady state of the system.

\begin{figure}
\centering
\includegraphics[width= \columnwidth]{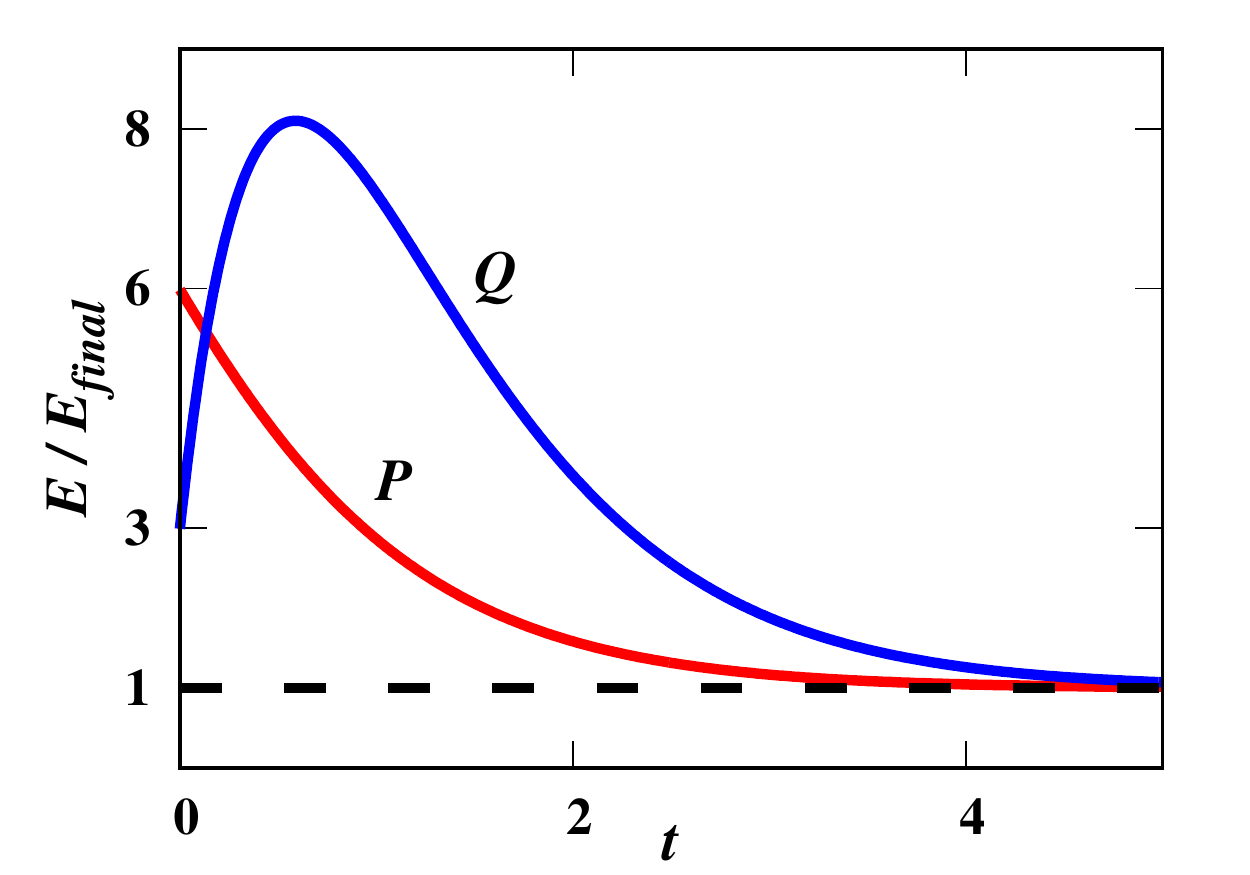} 
\caption{\label{figure1 monodispersed}The time evolution of the mean kinetic energy, $E$ of the  mono-dispersed Maxwell gas for two systems $P$ and $Q$ with initial conditions  $E^P(0)=4$, $E^Q(0)=2$, $C^P(0)=3$ and $C^Q(0)=23$, which satisfy the conditions for the Mpemba effect as described in Eq.~(\ref{condition mpemba monodispersed}).  The choice of the other parameters defining the system are $r=0.5$, $r_w=0.5$ and $\sigma=1$. $P$ relaxes to the steady state faster than $Q$, though its initial energy is larger. The time at which the trajectories cross each other is $\tau=0.1334$, as given by Eq.~(\ref{mpemba monodispersed}).}
\end{figure}

In Fig.~\ref{figure1 monodispersed}, we demonstrate the time evolution of the energies of two systems with initial conditions that satisfy Eq.~(\ref{condition mpemba monodispersed}). Though the initial energy of $P$ is larger, it relaxes to the steady state faster than $Q$.  As predicted by  Eq.~(\ref{mpemba monodispersed}), the two relaxation trajectories cross at some finite time.

\begin{figure}
\centering
\includegraphics[width=\columnwidth]{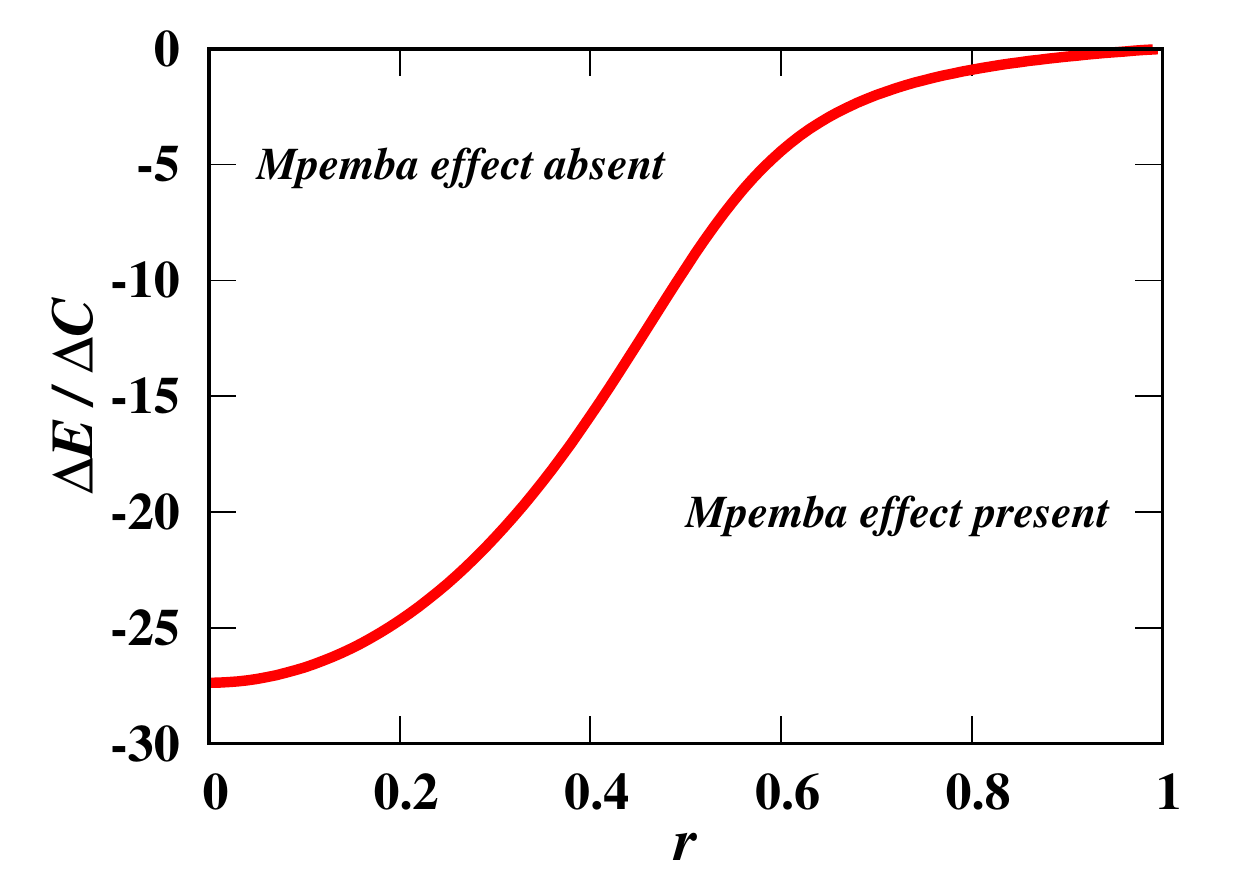}
\caption{The $\Delta E/\Delta C$--$r$ phase diagram showing regions where  the Mpemba effect is observed for mono-dispersed Maxwell gas (see Sec.~\ref{sec: mono-dispersed gas}), where $r$ is the coefficient of restitution. All other parameters are kept constant.  The region below the critical line show   the Mpemba effect.}\label{fig01a}
\end{figure}

Keeping all other parameters fixed, and allowing only the coefficient of restitution to vary, we can identify the region of phase space (initial condition) where   the Mpemba effect is observable, based on Eq.~(\ref{condition mpemba monodispersed}).  This is shown in Fig.~\ref{fig01a}. Clearly, as $r$ decreases to zero, the correlations need to be large for   the Mpemba effect to be present.
On the other hand, that in the near elastic limit $r\rightarrow 1$, it is much easier to observe the Mpemba effect, as the only requirement is that   $\Delta C$ and $\Delta E$ are anti-correlated. However, we note that even in this case, non-zero correlations imply that the system is not in a steady state.   

\section{The Mpemba effect in bi-dispersed Maxwell gas \label{bidispersed}}

In Sec.~\ref{sec: mono-dispersed gas}, we discussed the possibility of  the Mpemba effect in mono-dispersed gas.  The Mpemba effect was only present when the initial states where different from the steady states at that corresponding temperature. We now generalize the analysis to bi-dispersed gases,  based on an analysis of Eq.~(\ref{time-ev}), and show the presence of   the Mpemba effect even when the initial states are restricted to steady states.

In a bi-dispersed gas, the temperatures of the two components  are generally different [see Appendix~\ref{appendix a bidispersed}]. We denote them by $E_A$ and $E_B$. We denote the total kinetic energy of the system by $E_{tot}$, where 
\begin{align}
E_{tot}=E_A+E_B,
\label{E1 define}
\end{align}
and the difference in energies by $E_{diff}$:
\begin{align}
E_{diff}=E_A-E_B.
\label{E2 define}
\end{align}

We  define   the Mpemba effect in bi-dispersed gases similarly to the definition in the mono-dispersed gases. To this end, we consider two systems $P$ and $Q$ where $E_{tot}$ of $P$ is larger. Both $P$ and $Q$ are initially in their steady states. We then quench both systems to a lower temperature.   The Mpemba effect is present in this case when the $E_{tot}$ trajectories of the two systems cross each other.

We next consider separately the cases of one or both components driven. The reason for this separation is that in some experiments only one component is driven~\cite{Baxter:03,baxter2007temperature,Baxter-PRL:2007,Comb-PRE:2008,Burdeau-PRE:2009,Yule:2013}, whereas in others both components are driven~\cite{Feitosa2002,barrat2002lack,Pagnani-PRE:2002,Wang2008,JJBrey-non-eq:2009,Uecker:2009}. The respective analysis may be found in Secs.~\ref{sec: one component driven} and \ref{sec: both component driven}.

\subsection{One component is driven \label{sec: one component driven}}

Consider a bi-dispersed driven Maxwell gas where only  component $A$ is driven with driving strength $\sigma$. 
The time evolution equation [see Eq.~(\ref{time-ev})] for the correlation functions can be expressed in terms of $E_{tot}$ and $E_{diff}$ as
\begin{equation}
\frac{d\boldsymbol{E}(t)}{dt}=-~\boldsymbol{\chi}\boldsymbol{E}(t)+\boldsymbol{D}, \label{time ev E}
\end{equation}
where
\begin{align}
\boldsymbol{E}(t)&=\begin{bmatrix}
E_{tot}(t), &
E_{diff}(t)
\end{bmatrix}^T, \\
\boldsymbol{D}&=\begin{bmatrix}
\lambda_d \sigma^2, &
\lambda_d \sigma^2
\end{bmatrix}^T,
\end{align}
and $\boldsymbol{\chi}$ is a $2\times2$ matrix with components
$\chi_{11}$, $\chi_{12}$, $\chi_{21}$ and $\chi_{22}$  as given in Eq.~(\ref{appendix : chi}). The details of the calculation are shown in Appendix~\ref{appendix bidispersed}. Equation~(\ref{time ev E}) can be solved exactly by linear decomposition using the eigenvalues $\lambda_\pm$ of $\boldsymbol{\chi}$:
\begin{equation}
\lambda_\pm=\frac{1}{2}\Big[(\chi_{11}+\chi_{22})\pm \sqrt{(\chi_{11}-\chi_{22})^2+4\chi_{12}\chi_{21}}\Big].
\end{equation}

It is straightforward to show that $\lambda_{\pm}>0$ with $\lambda_+>\lambda_-$. The solution for $E_{tot}(t)$ and $E_{diff}(t)$ is
\begin{align}
\begin{split}
E_{tot}(t)&=K_+e^{-\lambda_+t}+ K_-e^{-\lambda_-t}+K_0, \\
E_{diff}(t)&=L_+e^{-\lambda_+t}+ L_-e^{-\lambda_-t}+L_0,
\end{split} \label{sol E1 one driven}
\end{align}
where the coefficients $K_+, K_-,K_0, L_+, L_-$ and $L_0$ are as given in Eq.~(\ref{appendix coeff one component driven}).

We now consider two  systems labeled as $P$ and $Q$ with different initial conditions $(E^P_{tot}(0), E^P_{diff}(0))$ and $(E^Q_{tot}(0), E^Q_{diff}(0))$ where $E^P_{tot}(0)> E^Q_{tot}(0)$. Both the systems are quenched to a  common steady state whose total energy is smaller than the initial total energies of $P$ and $Q$. This is achieved when the systems $P$ and $Q$ are now driven with the same driving strength~($\sigma$) for the component $A$, while keeping all the other parameters same for both the systems.

 The Mpemba effect is present when  the  trajectories $E^P_{tot}(t)$ and $E^Q_{tot}(t)$ cross each other at some finite time $t=\tau$ at which
\begin{equation}
E^P_{tot}(\tau)=E^Q_{tot}(\tau).
\end{equation}
Substituting into Eq.~(\ref{sol E1 one driven}), we obtain 
\begin{equation}
K^P_+e^{-\lambda_+\tau}+ K^P_-e^{-\lambda_-\tau}=K^Q_+e^{-\lambda_+\tau}+ K^Q_-e^{-\lambda_-\tau}, \label{evolution eq at crossing}
\end{equation}
whose solution is
\begin{equation}
\tau=\frac{1}{\lambda_+-\lambda_-}\ln \Big[\frac{K^P_+-K^Q_+}{K^Q_--K^P_-} \Big],
\end{equation}
which in terms of the initial conditions reduce to
\begin{equation}
\tau=\frac{1}{\lambda_+-\lambda_-} \ln \Big[\frac{\chi_{12}\Delta E_{diff}-(\lambda_--\chi_{11})\Delta E_{tot}}{\chi_{12}\Delta E_{diff}-(\lambda_+-\chi_{11})\Delta E_{tot}}  \Big], \label{crossing time one driven}
\end{equation}
where
\begin{align}
\begin{split}
&\Delta E_{tot}=E^P_{tot}(0)-E^Q_{tot}(0), \\
&\Delta E_{diff}=E^P_{diff}(0)-E^Q_{diff}(0).
\end{split} \label{eq: delta E}
\end{align}
Following the same argument as for the case of mono-dispersed Maxwell gas in Sec.~\ref{sec: mono-dispersed gas}, Eq.~(\ref{crossing time one driven}) leads to the criterion for the crossing of the two trajectories as
\begin{equation}
\chi_{12}\Delta E_{diff}>(\lambda_+-\chi_{11})\Delta E_{tot}. \label{condition one driven}
\end{equation}

\begin{figure}
\centering
\includegraphics[width=\columnwidth]{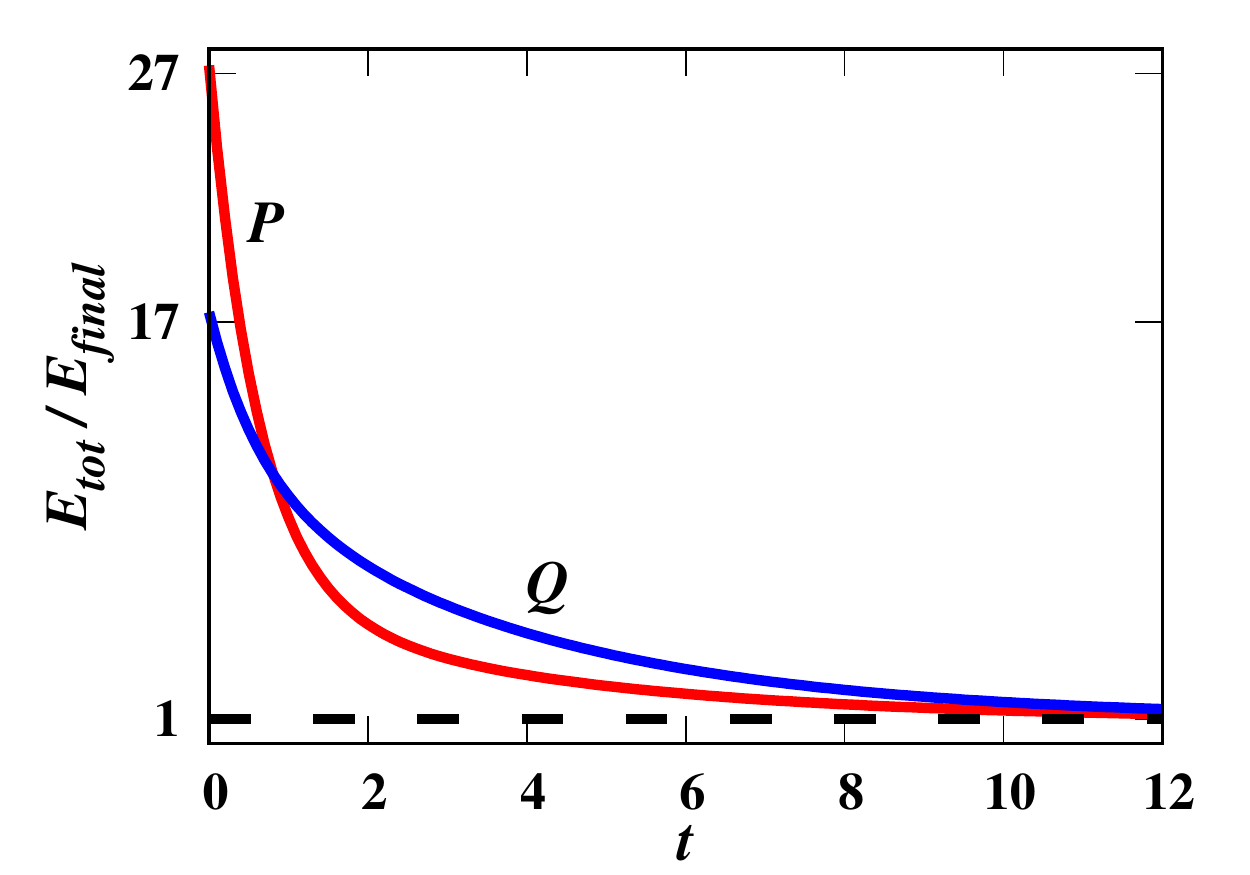}
\caption{The time evolution of the total energy, $E_{tot}(t)$  for two systems $P$ and $Q$  of the bi-dispersed Maxwell gas where only one component is driven with initial conditions: $E^P_{tot}(0)=22$, $E^Q_{tot}(0)=14$, $E^P_{diff}(0)=18$ and $E^Q_{diff}(0)=4$ such that $E^P_{tot}(0)>E^Q_{tot}(0)$, which satisfies the condition for the Mpemba effect as described in Eq.~(\ref{condition one driven}).  The choice of the other parameters defining the system are $m_B=2 m_A$, $r_{AA}=r_{AB}=r_{BB}=r_w=0.5$, $\nu_A=0.2$, $\nu_B=0.8$ and $\sigma=1$. $P$ relaxes to the steady state faster than $Q$, though its initial energy is larger. The time at which the trajectories cross each other is $\tau$=0.807 as given by Eq.~(\ref{crossing time one driven}).}\label{figure02}
\end{figure}

In Fig.~\ref{figure02}, we consider such a situation where Eq.~(\ref{condition one driven}) is satisfied. The trajectories cross at the point as predicted by Eq.~(\ref{crossing time one driven}). It is clear that though $P$ has larger initial energy, it relaxes faster.

\begin{figure}
\centering
\includegraphics[width=\columnwidth]{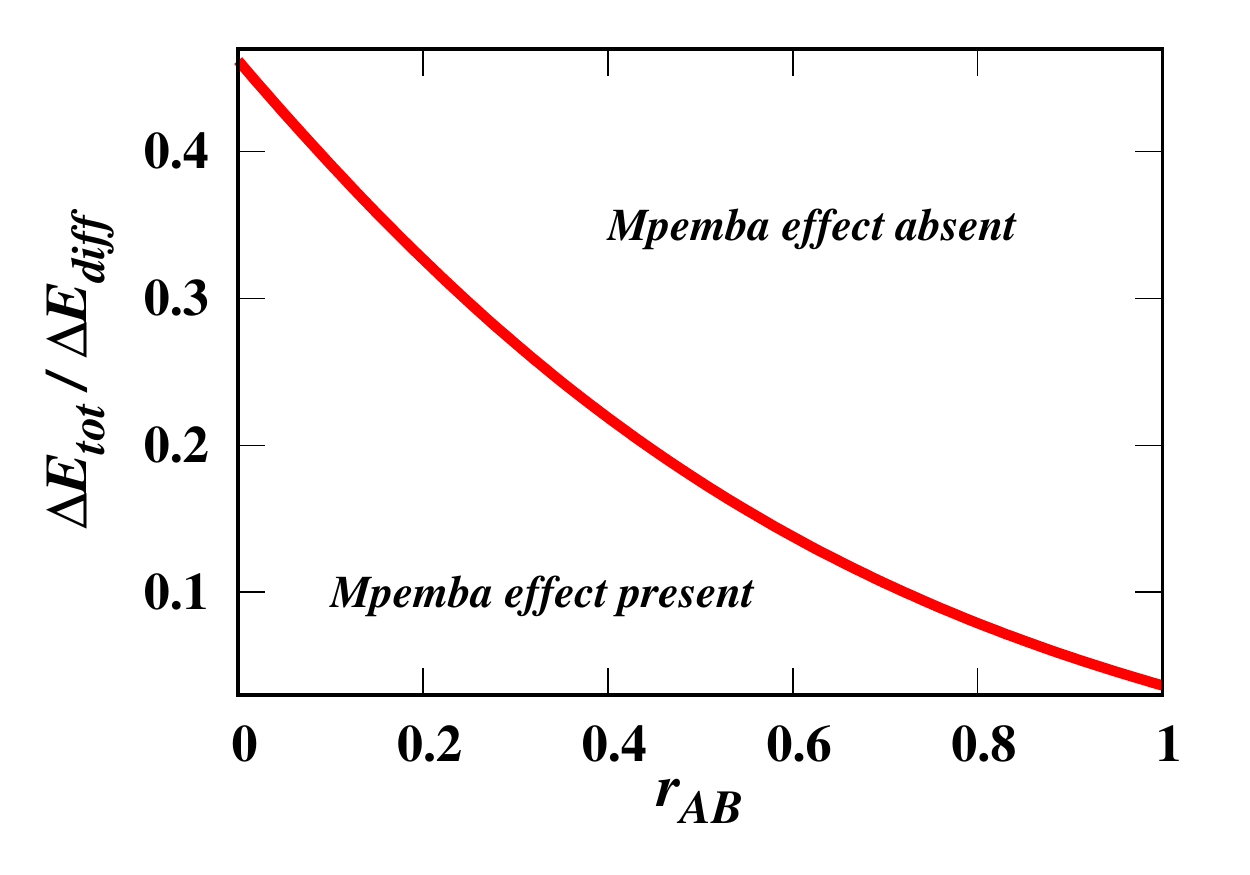}
\caption{The $\Delta E_{tot}/\Delta E_{diff}$--$r_{AB}$ phase diagram showing regions where   the Mpemba effect is observed for the bi-dispersed Maxwell gas, where only one component of the gas is driven (see Sec.~\ref{sec: one component driven}) and $r_{AB}$ is the coefficient of restitution. All other parameters are kept constant.  The choice of the other parameters defining the system are $m_A=2m_B$, $\nu_A=\nu_B=0.5$ and  $r_{AA}=r_{BB}=r_w=0.5$. The region below the critical line shows   the Mpemba effect whereas the region on the other side of the critical line does not show  the Mpemba effect.}\label{fig03}
\end{figure}

In Fig.~\ref{fig03}, we identify the region of phase space (initial condition) where   the Mpemba effect is observable, based on Eq.~(\ref{condition one driven}). In the figure, the variation with $r_{AB}$ is shown. The region below the line in the phase diagram shows   the Mpemba effect whereas the other region does not show the effect.
 
In the above analysis, the systems $P$ and $Q$ have the same parameters once the quench is done. However, in the initial states, the parameters -- reaction rates, coefficients of restitution, driving strength -- could be different for $P$ and $Q$. These parameters, though not explicitly mentioned, enter through the initial values $E_{tot}$ and $E_{diff}$.  As a result, one can tune the parameters appropriately to obtain initial steady states that satisfy the condition given by Eq.~(\ref{condition one driven}) and hence show   the Mpemba effect. 

We now ask a more refined question. Let us suppose that the systems $P$ and $Q$ have the same parameters throughout (both initially, as well as after the quench) except for the driving strength, which is different initially and the same after the quench. Can   the Mpemba effect be present in this case, when only component $A$ is driven?
The condition for   the Mpemba effect to be present is the same as that derived for the more general case [see Eq.~(\ref{condition one driven})]. However, when all parameters other than driving strength are kept identical, the ratio $\Delta E_{tot}/\Delta E_{diff}$  in the initial steady states has a simple form:
\begin{equation}
\frac{\Delta E_{tot}}{\Delta E_{diff}}=\frac{1+\lambda_{AB} \nu_A X^2_A \mathcal{Q}}{1-\lambda_{AB} \nu_A X^2_A \mathcal{Q}}>1, \label{steady state condition mpemba}
\end{equation}
where $ \mathcal{Q}$ is defined in equation~(\ref{eq:Q}).  Note that $ \mathcal{Q}>0$ and hence the ratio in Eq.~(\ref{steady state condition mpemba}) is always larger than one. On the other hand, the ratio $\Delta E_{tot}/\Delta E_{diff}$ should always be less than 
$\chi_{12}/(\lambda_{+}-\chi_{11})$ for one to observe   the Mpemba effect and it can be shown that the maximum value of the quantity is one. 
Thus, Eq.~(\ref{steady state condition mpemba}) does not satisfy the required condition for the existence of  a Mpemba effect.

So far we have discussed the possibility of  having the Mpemba effect in a bi-dispersed Maxwell gas where only one component is driven. We showed that, as compared to the mono-dispersed gas, for initial states that correspond to steady states, the bi-dispersed gas shows   a Mpemba effect for a wide range of parameters. However, when the two systems $P$ and $Q$ are identical except for the driving strength,   the Mpemba effect is not possible for steady state initial conditions. We now generalize the calculations to a bi-dispersed gas where both components are driven, and show that even for systems that differ only by the driving strength,   the Mpemba effect can be observed.

\subsection{Both components are driven \label{sec: both component driven}}

Next we consider a bi-dispersed Maxwell gas where both the components of the gas are driven. Here, type $A$ and $B$ particles of the bi-dispersed Maxwell gas are driven at the same rate $\lambda_d$ but with different driving strengths $\sigma_A$ and $\sigma_B$, respectively. The time evolution of the quantities $E_{tot}$ and $E_{diff}$ are given by Eq.~(\ref{time ev E}) with the column matrix $\boldsymbol{D}$ of the form
\begin{align}
\boldsymbol{D}&=\begin{bmatrix}
\lambda_d (\sigma^2_A+\sigma^2_B), &
\lambda_d (\sigma^2_A-\sigma^2_B)
\end{bmatrix}^T.
\end{align}
The solutions for $E_{tot}(t)$ and $E_{diff}(t)$ are obtained in a similar way as in Eq.~(\ref{sol E1 one driven}), but the coefficients $K_+, K_-,K_0, L_+, L_-$ and $L_0$ are now given by Eq.~(\ref{appendix coeff both components driven}).

Our main aim is to look for the existence of   a Mpemba effect by considering two systems with identical parameters, except for the driving strengths.  To this end, we consider two such similar systems, $P$ and $Q$, driven with different noise strengths, thus attaining different initial steady states with different initial total energies.

\begin{figure}
\centering
\includegraphics[width=\columnwidth]{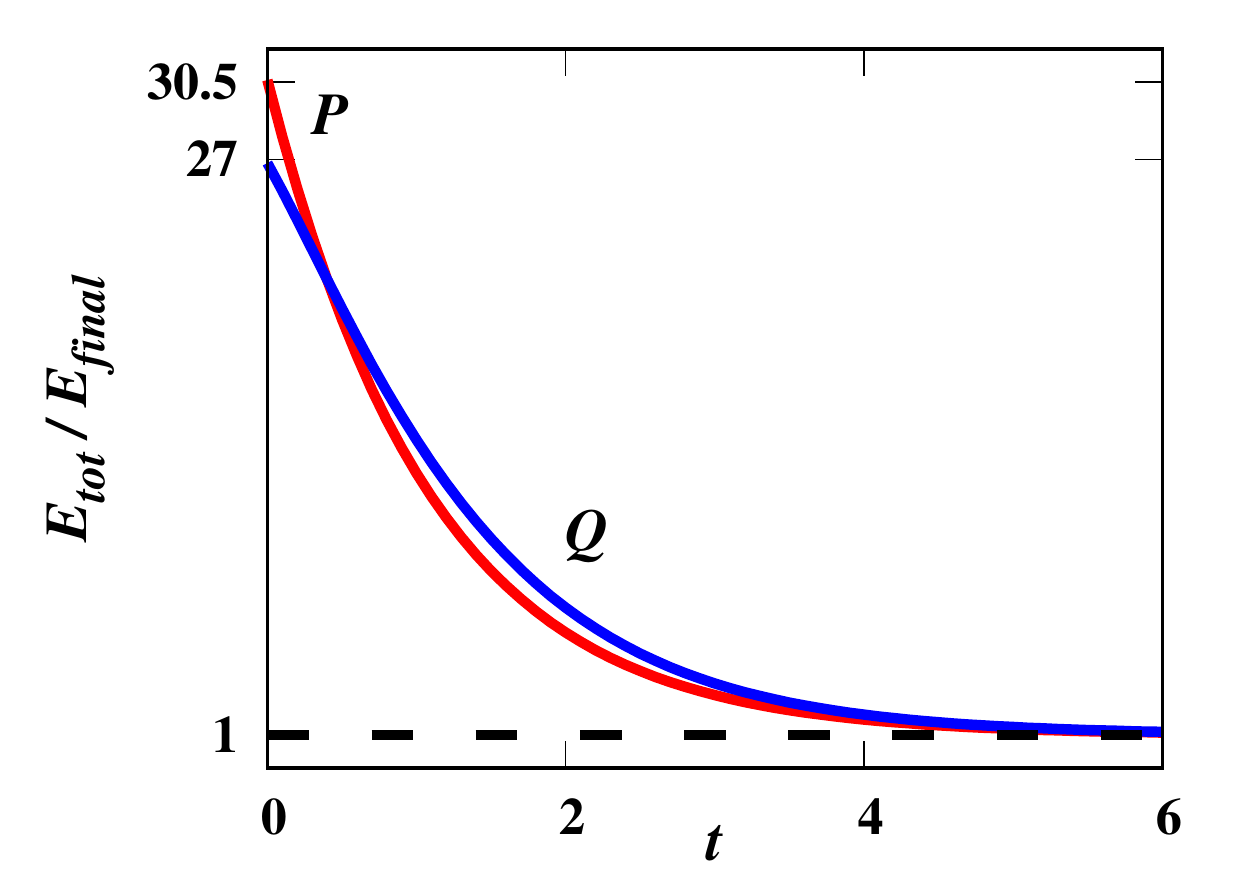}
\caption{The time evolution of the total energy, $E_{tot}(t)$ for two systems $P$ and $Q$  of a bi-dispersed Maxwell gas where both components are driven with initial steady state conditions: $E^P_{tot}(0)=150.4$, $E^Q_{tot}(0)=132$, $E^P_{diff}(0)=75.2$ and $E^Q_{diff}(0)=30.5$ such that $E^P_{tot}(0)>E^Q_{tot}(0)$, which satisfies the condition for the Mpemba effect as described in Eq.~(\ref{condition one driven}).  The choice of the other parameters defining the system are $m_B=10 m_A$, $r_{AA}=r_{BB}=r_w=0.5$, $r_{AB}=0.6$, $\nu_A=0.2$, $\nu_B=0.8$, $\sigma_A=2$ and $\sigma_B=1$. $P$ relaxes to the steady state faster than $Q$, though its initial energy is larger. The time at which the trajectories cross each other is $\tau$=0.39 as given by Eq.~(\ref{crossing time one driven}).}
\label{fig05}
\end{figure}

Let system $P$ have higher initial total energy compared to $Q$. Both  systems are then driven to a common steady state with a lower energy compared to the initial steady state of systems $P$ and $Q$. This is achieved when  $P$ and $Q$ are driven with the same driving strengths ($\sigma_A$ and $\sigma_B$) for the components $A$ and $B$ .
The cross-over time $\tau$ for the crossing of the trajectories of $E^P_{tot}(t)$ and $E^Q_{tot}(t)$ is obtained using Eq.~(\ref{crossing time one driven}), and the criterion for the occurrence of   the Mpemba effect is given by Eq.~(\ref{condition one driven}). An example of such a crossing is shown in Fig.~\ref{fig05}. 
%However, the ratio $\Delta E_{tot}/\Delta E_{diff}$ depends only on the initial steady state energies of the systems as can be seen from Eq.~(\ref{eq: delta E}) and has a form given by Eq.~(\ref{critical condition ME both driven}). These are the accessible steady state values for the ratio $\Delta E_{tot}/\Delta E_{diff}$. 

\begin{figure}
\centering
\includegraphics[width=\columnwidth]{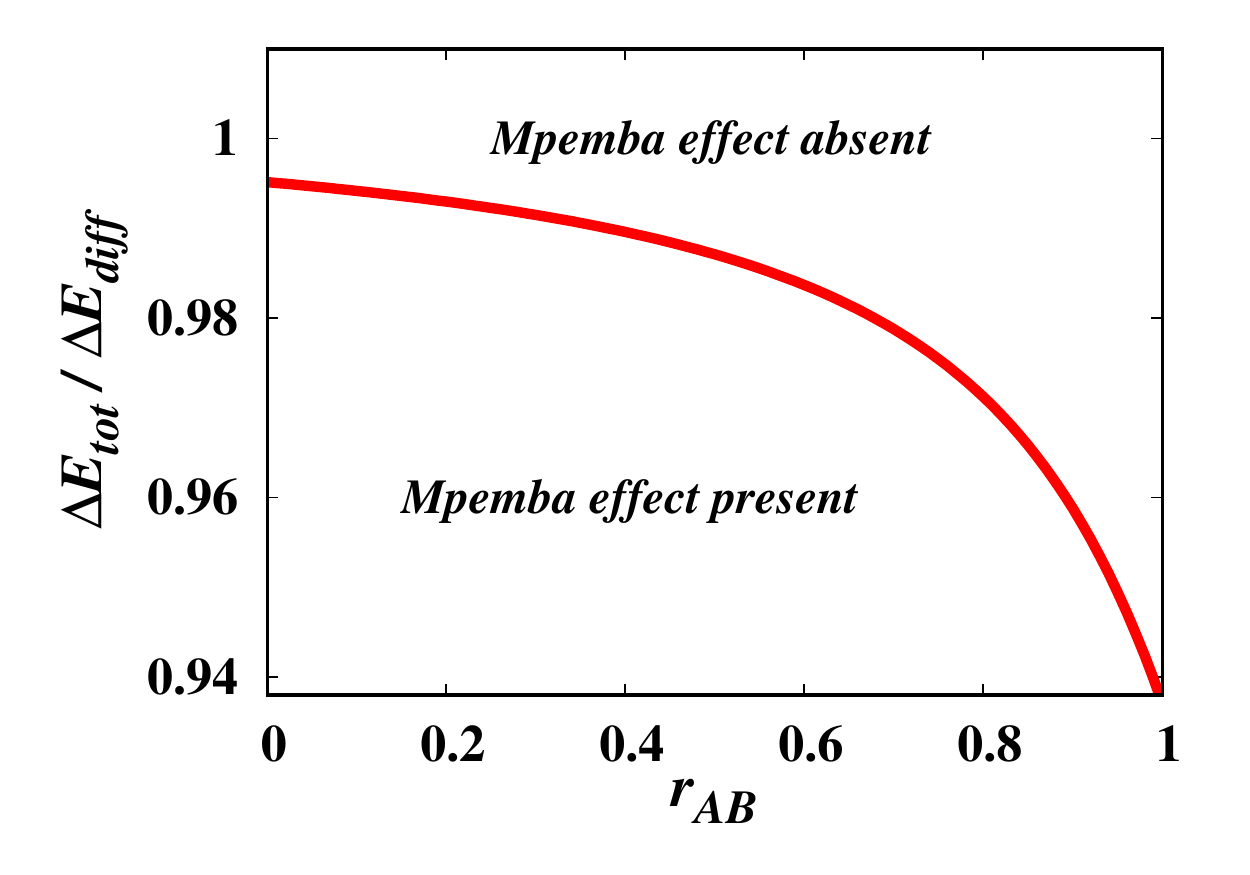}
\caption{The $\Delta E_{tot}/\Delta E_{diff}$--$r_{AB}$ phase diagram showing regions where   the Mpemba effect is observed for the bi-dispersed Maxwell gas, where both components of the gas are driven (see Sec~\ref{sec: both component driven}) and $r_{AB}$ is the coefficient of restitution. All other parameters are kept constant.  The choice of the other parameters defining the system are $m_B=2m_A$, $\nu_A=0.2$, $\nu_B=0.8$, $r_w=0.6$ and  $r_{AA}=r_{BB}=0.5$. The region below the critical line gives the set of initial steady states that show  the Mpemba effect whereas the region on the other side of the critical line correspond to initial states that do not show the Mpemba effect.   }
\label{fig06}
\end{figure}

In Fig.~{\ref{fig06}}, based on Eq.~(\ref{condition one driven}), we identify the region of phase space (initial condition) where the Mpemba effect is observable. In the figure, the variation with $r_{AB}$ is shown. If the ratio $\Delta E_{tot}/\Delta E_{diff}$ falls in the region below (above) the line in the phase diagram [see Fig.~{\ref{fig06}}], then the system exhibits (does not exhibit) the Mpemba effect. For steady state initial conditions, the ratio $\Delta E_{tot}/\Delta E_{diff}$  depends on parameters as  given by Eq.~(\ref{critical condition ME both driven}). Note that the ratio $\Delta E_{tot}/\Delta E_{diff}$ is also a function of the driving strengths, $\sigma_A$ and $\sigma_B$ [see Eq.~(\ref{critical condition ME both driven})]. It turns out that   the driving strengths can be appropriately tuned to access the entire region of phase space (initial condition) in which  the Mpemba effect is observable.

\subsection{The Inverse Mpemba Effect \label{sec: inverse mpemba}}

\begin{figure}
\centering
\includegraphics[width=\columnwidth]{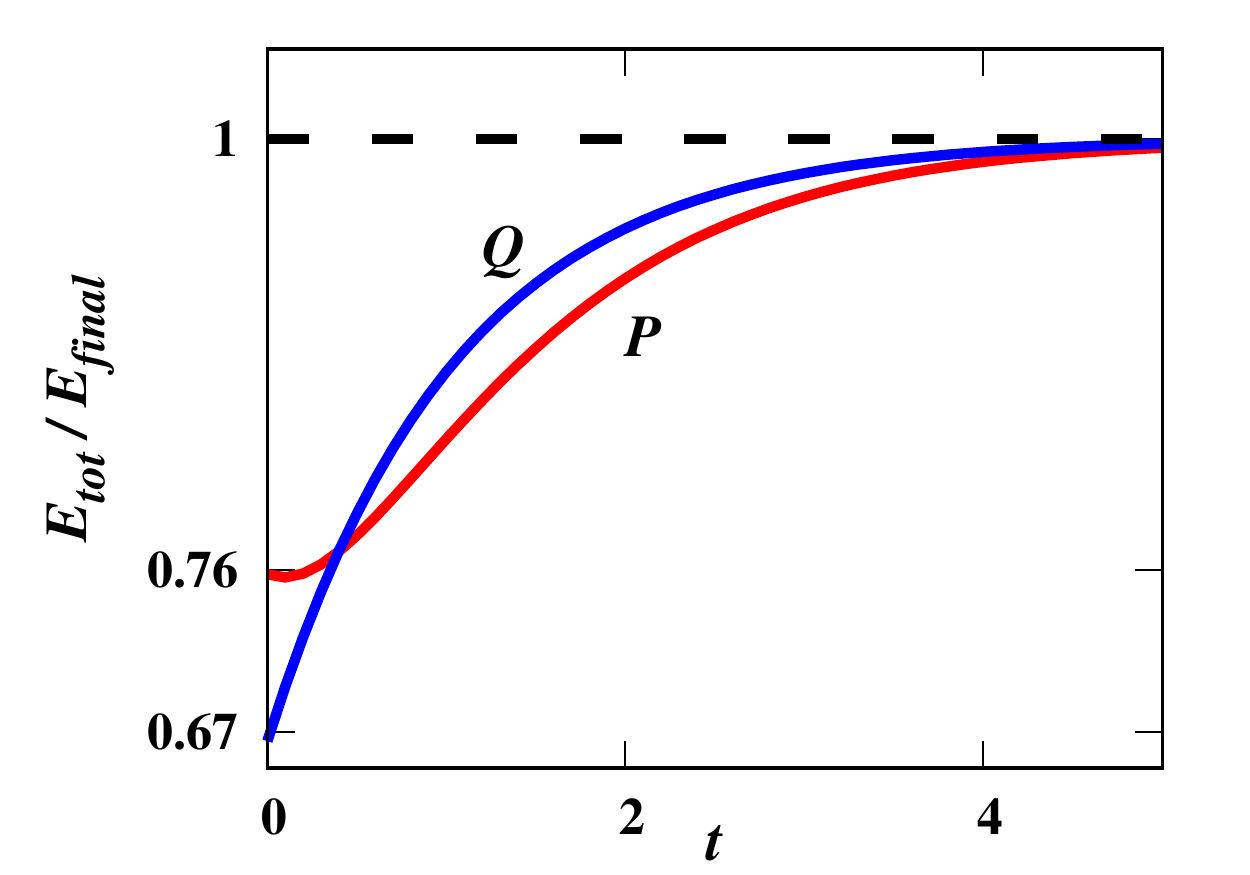} 
\caption{The time evolution of the total energy, $E_{tot}(t)$ for two systems $P$ and $Q$ of bi-dispersed Maxwell gas where both components are driven  with initial steady state conditions: $E^P_{tot}(0)=150.4$, $E^Q_{tot}(0)=132$, $E^P_{diff}(0)=75.2$ and $E^Q_{diff}(0)=30.5$ such that $E^P_{tot}(0)>E^Q_{tot}(0)$, which satisfies the condition for the inverse Mpemba effect as described in Eq.~(\ref{condition one driven}).   The choice of the other parameters defining the system are $m_B=10 m_A$, $r_{AA}=r_{BB}=r_w=0.5$, $r_{AB}=0.6$, $\nu_A=0.2$, $\nu_B=0.8$, $\sigma_A=8$ and $\sigma_B=8$. $P$ relaxes to the steady state slower than $Q$, though its initial energy is larger. The time at which the trajectories cross each other is $\tau$=0.39 as given by Eq.~(\ref{crossing time one driven}).} \label{inverse mpemba}
\end{figure}

 The inverse Mpemba effect refers to the phenomenon that, when quenched to a high temperature,  an initially colder system heats faster than a system at intermediate temperature. The analysis for showing   the Mpemba effect in Sec.~\ref{sec: both component driven} can be generalized to show   the inverse Mpemba effect in the driven binary gas when both components are driven. We prepare two  systems $P$ and $Q$ in steady states such that the total energy of $P$ is larger than that of $Q$.  Both  systems are then quenched, using suitable driving strengths for the individual components of the bi-dispersed gas, to a common steady state having a higher energy compared to the initial energies of both $P$ and $Q$. The cross-over time $\tau$ for the crossing of the trajectories of $E^P_{tot}(t)$ and $E^Q_{tot}(t)$ is obtained using Eq.~(\ref{crossing time one driven}) and the criterion for the occurrence of  the inverse Mpemba effect is given by Eq.~(\ref{condition one driven}). 
An example of such a crossing is shown in  Fig.~\ref{inverse mpemba}.  

The accessible steady states of the system that satisfy the condition for   the inverse Mpemba effect turns out to be the same as that for   the direct Mpemba effect and can be obtained using Eq.~(\ref{critical condition ME both driven}). Thus Fig.~\ref{fig06} also illustrates that the valid steady states of the system given by Eq.~(\ref{critical condition ME both driven}) belongs to the region of phase space (initial condition) given by Eq.~(\ref{condition one driven}) where   the inverse Mpemba effect is observable.

\subsection{The Strong Mpemba Effect}
We now explore the possibility of   a strong Mpemba effect in the binary Maxwell gas. The strong Mpemba effect refers to the phenomenon wherein the system at higher temperature relaxes to a final steady state exponentially faster, namely with a larger exponential rate compared to other initial conditions. Up to now, we had only considered the case where the trajectories cross, which in general does not imply that the decay rate at large times is different. The linear evolution equation in  Eq~(\ref{time-ev})  allows certain set of initial conditions to relax to the final steady state exponentially faster compared to other initial states. The effect may be realized when the coefficient ($K_-$) associated with the slower relaxation rate  in the time evolution of total kinetic energy, $E_{tot}(t)$ [see Eq.~\ref{sol E1 one driven}] vanishes.  In what follows, we would like to probe the system of bi-dispersed gas with both type of particles driven to look for  the presence of the strong Mpemba effect.

Setting the coefficient $K_-$ [given by Eq.~(\ref{appendix coeff both components driven})]  to zero, we obtain
\begin{align}
E_{tot}(0)=\frac{\chi_{12}}{\lambda_+ - \chi_{11}}E_{diff}(0)-c, \label{strong mpemba}
\end{align}
where
\begin{align}
c=\frac{\lambda_d \left[\left( \chi_{12}-\lambda_++\chi_{11} \right)\sigma^2_A - \left(\chi_{12}+\lambda_+-\chi_{11}  \right) \sigma^2_B \right]}{\lambda_- (\lambda_+ - \chi_{11})}.
\end{align}
For a system with all other parameters kept fixed, solution of Eq.~(\ref{strong mpemba}) in terms of 
$E_{tot}(0)$ and $E_{diff}(0)$  provides the set of initial states whose relaxation is 
exponentially faster than the  set of generic states.  Note that the set of initial states that satisfy Eq.~(\ref{strong mpemba}) lie on a straight line.

Among these initial states one would like to determine the ones which are steady states. Remember that
the steady state ratio of $E_{tot}(0)/E_{diff}(0)$ for a system is given by
\begin{align}
\frac{E_{tot}(0)}{E_{diff}(0)}=f(\sigma_A,\sigma_B),
\end{align}
where $f(\sigma_A,\sigma_B)$ is given by Eq.~(\ref{appendix: steady state ratio energies}) and is only a function of driving strengths ($\sigma_A$ and $\sigma_B$) as all other parameters are kept constant. 
One observes that valid steady states with initial energies, $E_{tot}(0)$ and $E_{diff}(0)$ that satisfy the condition for the strong Mpemba effect [see Eq.~(\ref{strong mpemba})] can be obtained by appropriately tuning the driving strengths $\sigma_A$ and $\sigma_B$.

\begin{figure}
\centering
\includegraphics[width=\columnwidth]{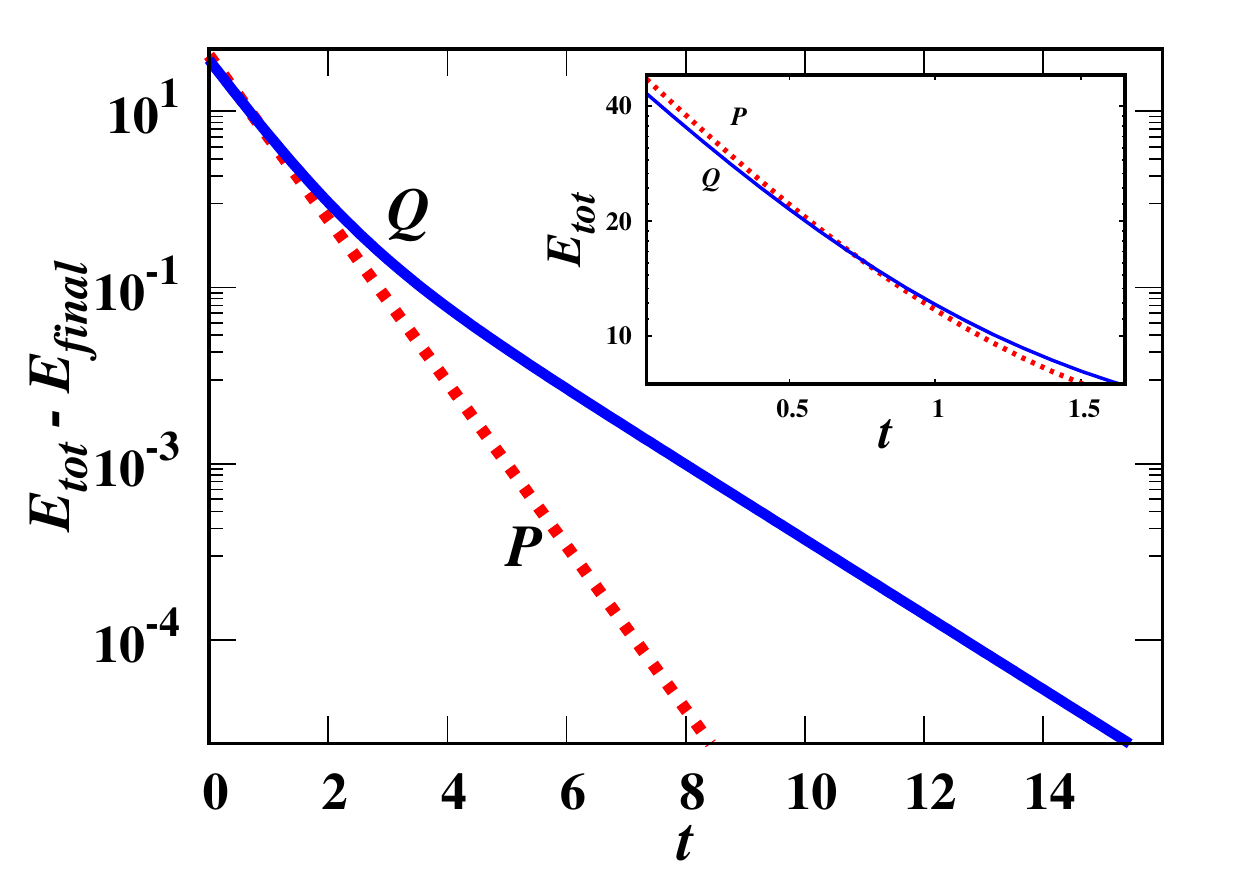} 
\caption{The time evolution of the total energy, $E_{tot}(t)$ for two systems $P$ and $Q$ of bi-dispersed Maxwell gas, both components are initially in steady states, with  $E^P_{tot}(0)=47.8$, $E^Q_{tot}(0)=43.8$, $E^P_{diff}(0)=45.6$ and $E^Q_{diff}(0)=39$ such that $E^P_{tot}(0)>E^Q_{tot}(0)$. These initial values satisfy both  the conditions for  the Mpemba effect as described in Eq.~(\ref{condition one driven}) as well as those  for   the strong Mpemba effect (for system $P$) as described in Eq.~(\ref{strong mpemba}). The choice of the other parameters defining the system are $m_B=10 m_A$, $r_{AA}=0.5$, $r_{BB}=0.4$, $r_w=0.6$, $r_{AB}=0.35$, $\nu_A=0.2$, $\nu_B=0.8$, $\sigma_A=2$ and $\sigma_B=1$. $P$ equilibrates  to the final state at an exponentially faster rate compared to $Q$. Inset: The trajectories for $P$ and $Q$ cross at a time $\tau=0.73$,   as given by Eq.~(\ref{crossing time one driven}). } \label{fig08}
\end{figure}

Thus for a system of bi-dispersed Maxwell gas where both components are driven, there exists steady state initial conditions
which satisfy the condition  given by Eq.~(\ref{strong mpemba}) and hence approach the final steady state exponentially faster compared to  any other  similar system whose initial energies lie slightly below or above the line. An example of the strong Mpemba effect is shown  in Fig.~\ref{fig08}. The figure shows the evolution of the total energy $E_{tot}$ as a function of time $t$ for two bi-dispersed systems  $P$ and $Q$ that have identical parameters except for the initial driving strength after being quenched to a lower temperature by changing the driving strength to a common lower value.  The initial state of the system $P$  is chosen in such a way that $[E_{tot}(0),E_{diff}(0)]$ satisfies the condition for the strong Mpemba effect [Eq.~(\ref{strong mpemba})], and hence evolves to the final state with a single faster relaxation rate. On the other hand, in system $Q$,  the initial state does
not satisfy the strong Mpemba effect condition, thus it relaxes differently, and asymptotically evolves with the slower rate.
Further, as the initial condition of the two systems $P$ and $Q$ happen to satisfy the relation for the existence of  the Mpemba effect  [Eq.~(\ref{condition one driven})], the trajectory of the system $P$ with higher initial energy crosses that of $Q$ with lower initial energy.  The crossing time $\tau$ could be obtained using Eq.~(\ref{crossing time one driven}) which is captured in the inset of the figure.

\section{Conclusion \label{conclusion}}

In summary, through an exact analysis of the driven mono-dispersed and bi-dispersed Maxwell gases, we derived the conditions under which the Mpemba effect, the inverse Mpemba and the strong Mpemba effect can be observed. In Maxwell gas, the rate of collision between particles is independent of the relative velocity. In addition, the  well-mixed limit is assumed such that spatial correlations are ignored. The equations for the two point correlations close among themselves resulting in a coupled set of linear equations allowing for an exact solution. This linearity happens to be natural to the model and thus does not require any approximations  that have been employed in models where the collision rates are velocity dependent. To demonstrate the existence of  the Mpemba effect, we determine the conditions under which a hotter system relaxes faster than a cooler system when quenched to a temperature lower than both. For the case of mono-dispersed Maxwell gas, we showed that   the Mpemba effect is possible only if the initial states do not correspond to steady states. On the other hand, for bi-dispersed Maxwell gas, there is a range of parameters for which  the Mpemba effect exists, even when the states from which the quench is done are  restricted to valid steady states.   In a similar framework, we also demonstrated the existence of  the inverse Mpemba effect where a system is heated instead of cooled, i.e.,  a system at a lower initial temperature relaxes to a high temperature state faster than another system with an intermediate initial temperature. We also showed the existence of a stronger version of  the Mpemba effect, where a system equilibrates to a final steady state at an exponentially faster rate. 

 The exact analysis allows us to identify the reason behind the Mpemba effect in these systems. First, we note that, though the system evolves stochastically, the evolution of the history-averaged correlation functions is deterministic. The evolution equations are first  order differential equations, therefore two trajectories in the correlation functions-time phase space cannot intersect. However, for the mono-dispersed gas, the state of the system is defined by two quantities: energy and inter-particle correlation function. In this three dimensional space (third dimension being time), two trajectories cannot cross. Nevertheless, when projected onto the lower dimensional energy-time plane, trajectories may cross, leading to  the Mpemba effect. If the initial states correspond to steady states, then the inter-particle correlations are exactly zero and remain zero during time evolution, constraining  the correlation function-time plane to be two dimensional. Thus, a necessary condition  for  the Mpemba effect to be observed is a non-zero inter-particle correlations, or alternatively non-stationary initial states. On the other hand,
for bi-dispersed gases, there are two kinetic energies and three inter-particle correlation functions. Since the energy-time dimensions are themselves three dimensional, it is possible to observe  the Mpemba effect when the correlations are set to zero, as in a true steady state.
Note that for the present study, we have chosen intersection of total energy as an indicator of the Mpemba effect. This is a natural choice, as this is the quantity that is easiest to track in an experiment. It is possible to characterise the state of the system by more macroscopic quantities. For example, by  temperature $T$ and second Sonine coefficient, $a_2$ in the expansion of velocity distribution function around Gaussian, as studied for the case of granular systems of hard sphere particles~\cite{Lasanta-mpemba-1-2017}.

Our results are analogous to those found in the perturbative treatment of the most realistic granular systems~\cite{Lasanta-mpemba-1-2017,Torrente-rough-2019}. In these calculations,    the Mpemba effect can be seen in mono-dispersed systems when the initial conditions do not correspond to any steady state, but rather to some transient states that are close to the final steady state. This is achieved by choosing appropriate initial velocity distribution functions for the two systems. For the case of rough granular gas, calculations were carried out by considering states with Gaussian velocity distribution at all times which may not hold good for non-equilibrium systems. In contrast, the analysis in the present work does not make any assumption regarding the nature of velocity distribution in the steady state of the granular system. Therefore, extending the calculations of Refs.~\cite{Lasanta-mpemba-1-2017,Torrente-rough-2019} to smooth  bi-dispersed gas would be a promising area for future study. One can also search for exponentially faster relaxation protocols in these systems as studied in Refs.~\cite{PhysRevLett.124.060602,Klich-2019}. In addition, the results in this paper, particularly  the case of bi-dispersed Maxwell gas where both components are driven, suggest that driven binary gases are a good candidate for observation of  the Mpemba effect in granular experiments. 

Further, in the aim of demonstrating the existence of the Mpemba effect in Maxwell gases, we have assumed that energies evolve monotonically, by not explicitly accounting for the possibility that when a hot system is quenched to a lower temperature, the temperature could drop below the final temperature. In order to include scenarios with non-monotonic evolution of energy into the present framework, one may have to look at the behavior of the absolute values of the coefficients $K_-$~\cite{Lu-raz:2017}. 
This extended case may be addressed in future, to the present and other models of granular gases. 
One may also look at the relation between the existence of such non monotonic relaxation and the 
strong Mpemba effect, as the presence of non-monotonicity indicate a change in sign of the coefficient $K_-$.

\section{Acknowledgements}
O.R. is the incumbent of the Shlomo and Michla Tomarin career development chair, and is supported by the Abramson Family Center for Young Scientists and by the Israel Science Foundation, Grant No. 950/19.

\appendix

\section{ Time evolution for two point correlations \label{appendix two point correlation}}
In this appendix, we solve for the time evolution equations of two-point correlations as defined in sections \ref{mono-dispersed} and \ref{bi-dispersed} for the case of mono-dispersed and bi-dispersed Maxwell granular gas respectively.
\subsection{Mono-dispersed gas \label{appendix a monodispersed}}

For the mono-dispersed gas, the time evolution of mean kinetic energy, $E(t)$ and the two point velocity-velocity correlation function, $C(t)$ are given as in Eq.~(\ref{time evolution exact monodispersed}). The coefficients $K_+$, $K_-$, $K_0$, $L_+$, $L_-$ and $L_0$ in Eq.~(\ref{time evolution exact monodispersed}) are given by
\begin{widetext}
{\footnotesize{
\begin{align}
\begin{split}
K_+&=\frac{1}{\gamma}\left[(\lambda_--R_{11})\Sigma_1(0)-R_{12}\Sigma_2(0)-\frac{(\lambda_--R_{11})}{\lambda_+}\lambda_d\sigma^2\right], \qquad
K_-=\frac{1}{\gamma}\left[-(\lambda_+-R_{11})\Sigma_1(0)+R_{12}\Sigma_2(0)+\frac{(\lambda_+-R_{11})}{\lambda_-}\lambda_d\sigma^2\right],\\
K_0&=\frac{1}{\gamma}\left[\frac{\lambda_--R_{11}}{\lambda_+}-\frac{\lambda_+-R_{11}}{\lambda_-}\right]\lambda_d\sigma^2, \hspace{90pt} L_0=\frac{1}{\gamma}\left[ \frac{(\lambda_+-R_{11})(\lambda_--R_{11})}{R_{12}\lambda_+}-\frac{(\lambda_+-R_{11})(\lambda_--R_{11})}{R_{12}\lambda_-}\right]\lambda_d\sigma^2,\\
L_+&=\frac{1}{\gamma}\left[\frac{(\lambda_+-R_{11})(\lambda_--R_{11})}{R_{12}}\Sigma_1(0)-(\lambda_+-R_{11})\Sigma_2(0)-\frac{(\lambda_+-R_{11})(\lambda_--R_{11})}{R_{12}\lambda_+}\lambda_d\sigma^2\right],\\
L_-&=\frac{1}{\gamma}\left[\frac{-(\lambda_+-R_{11})(\lambda_--R_{11})}{R_{12}}\Sigma_1(0)+(\lambda_--R_{11})\Sigma_2(0)+\frac{(\lambda_+-R_{11})(\lambda_--R_{11})}{R_{12}\lambda_-}\lambda_d\sigma^2\right],\\
\gamma&=\lambda_+-\lambda_-. 
\end{split} \label{appendix coeff monodispersed}
\end{align}}}
\end{widetext}

\subsection{Bi-dispersed gas \label{appendix a bidispersed}}
For the case of bi-dispersed Maxwell gas, the two point correlations are defined as in Eq.~(\ref{2 point correlations}). First, we consider the general case where both components of the gas are driven. The results can be extended for one component driven bi-dispersed gas. We can write for the time evolution of the two point correlations using Eq.~(\ref{pa_single}) in a compact matrix form as
\begin{align}
\frac{d\boldsymbol{\Sigma}(t)}{dt}=~\boldsymbol{R}\boldsymbol{\Sigma}(t)+\boldsymbol{D},
\label{two-point-correlation-matrix-evolution}
\end{align}
where
\begin{align}
\boldsymbol{\Sigma}(t)&=\begin{bmatrix}
E_{A}(t), &
E_{B}(t), &
C_{AB}(t), &
C_{AA}(t), &
C_{BB}(t)
\end{bmatrix}^T, \\
\boldsymbol{D}&=\begin{bmatrix}
\lambda_d \sigma^2_A, &
\lambda_d \sigma^2_B, &
0, &
0, &
0
\end{bmatrix}^T,
\end{align}
where  $k$=$A, B$ and  the matrix $\boldsymbol{R}$ is given by
\begin{widetext}
{\footnotesize{
\begin{equation}
\boldsymbol{R}=\left[\begin{array}{ccccc} 
R_{2B}\!-\!R_{1A}\!-\!R_{3B}\!-\!R_d&R_{2B}&-2 R_{2B}+R_{3B}&2R_{1A}&0\\
R_{2A}&R_{2A}\!-\!R_{1B}\!-\!R_{3A}& -2 R_{2A}+R_{3A}&0&-R_{1B}\\
\frac{R_{3A}}{2 N_A}-R_4&\frac{R_{3B}}{2 N_B}-R_4&\frac{4R_4+R_{3B}-R_{3A}}{2}&-\frac{R_{3A}}{2 N_A}+\frac{R_{3A}}{2}  &-\frac{R_{3B}}{2 N_B}+\frac{R_{3B}}{2}\\
\frac{R_{1A}}{N_A-1}&0&R_{3B}& \frac{R_{1A}}{1-N_A}\!-\!R_{3B}\!-\!\frac{R_d}{1-r_w}&0\\
0&\frac{R_{1B}}{N_B-1}&R_{3B}&0&\frac{2R_{1B}}{N_B-1}\!-\!R_{3B}\end{array}\right].
\end{equation}}}
\end{widetext}
The constants  $R_{1k},R_{2k},R_{3k}, R_4, R_d$  are given by:
{\footnotesize 
\begin{align}
\begin{split}
R_{1k}&=\frac{\lambda_{kk}\alpha_{kk}(1-\alpha_{kk})(N_k-1)}{N} , \quad R_{2k}=\frac{\lambda_{AB}N_kX_k^2}{N},\\
R_{3k}&=\frac{2 R_{2k}}{X_k},  \quad R_4=\frac{\lambda_{AB}X_AX_B}{N},  \\
R_d &=\lambda_d(1-r_w^2).
\end{split}
\end{align}
}

In the steady state, in the thermodynamic limit, the inter-particle two point correlation functions $C_{ij}$, where $i,j \in (A,B)$ are zero, as shown in Ref.~\cite{Biswas_2020}. If in the initial state, these correlations are zero, then it remains zero for all times. We will be only considering such initial states. In that case, we can ignore these correlations, and write for the time evolution of mean kinetic energies $E_A$ and $E_B$ as 
{\small
\begin{widetext}
\begin{align}
\begin{split}
&\frac{dE_A(t)}{dt}=E_A(t)\Big(\lambda_{AB}\nu_B X^2_B-\lambda_{AA}\alpha_{AA}(1-\alpha_{AA})\nu_A - 2\lambda_{AB}\nu_B X_B-\lambda_d(1-r^2_w)\Big)
+E_B(t)\Big(\lambda_{AB}\nu_B X^2_B\Big)+\lambda_d \sigma^2_A \\
&\frac{dE_B(t)}{dt}=E_A(t)\Big(\lambda_{AB}\nu_A X^2_A\Big)
+E_B(t)\Big(\lambda_{AB}\nu_A X^2_A-\lambda_{BB}\alpha_{BB}(1-\alpha_{BB})\nu_B - 2\lambda_{AB}\nu_A X_A-\lambda_d(1-r^2_w)\Big)+\lambda_d \sigma^2_B \label{time sigma b}
\end{split}
\end{align}
\end{widetext}}
It can be written in a compact form as in Eq.~(\ref{time-ev}). In order to obtain the steady state values, we set the time derivative of Eq.~(\ref{time sigma b}) to zero. The steady state values for $E_A$ and $E_B$ are given by
{\small \begin{widetext} \begin{align}
&E_{A} = \frac{\begin{aligned} -\lambda_d \sigma^2_A\Big(-\lambda_d (1-r^2_w)-2\lambda_{AB}\nu_A X_A + \lambda_{AB} \nu_A X^2_A - \lambda_{BB} \nu_B (1-\alpha_{BB})\alpha_{BB}\Big) + \lambda_{AB} \lambda_d \nu_B X^2_B \sigma^2_B \end{aligned}}{\begin{aligned} 
\mathcal{F}
\end{aligned} }, \label{66}\\
&E_B = \frac{\begin{aligned} \lambda_{AB} \lambda_d \nu_A X^2_A \sigma^2_A-\lambda_d \sigma^2_B \Big(-\lambda_d(1-r^2_w)-2\lambda_{AB}\nu_B X_B + \lambda_{AB} \nu_B X^2_B - \lambda_{AA} \nu_A (1-\alpha_{AA})\alpha_{AA}\Big)  \end{aligned}}{\begin{aligned} 
\mathcal{F}
\end{aligned} } , \label{67}
\end{align}
where
\begin{align}
\mathcal{F}=-\lambda^2_{AB}\nu_A \nu_B X^2_A X^2_B + &\Big(-\lambda_d(1-r^2_w)-2\lambda_{AB}\nu_B X_B+\lambda_{AB}\nu_B X^2_B-\lambda_{AA}\nu_A(1-\alpha_{AA})\alpha_{AA} \Big) \nonumber \\ &\Big(-\lambda_d(1-r^2_w)-2\lambda_{AB}\nu_A X_A+\lambda_{AB}\nu_A X^2_A-\lambda_{BB}\nu_B(1-\alpha_{BB})\alpha_{BB} \Big). \label{eq:F}
\end{align}
\end{widetext}
}
One can do similar calculations to solve for the steady state mean kinetic energies, for the case where only one component (say $A$) is driven with driving strength ($\sigma$) and at a rate ($\lambda_d$). In that case, the mean kinetic energies of the components are given by

{\small  \begin{widetext} \begin{align}
&E_{A} = \frac{\lambda_d \sigma^2}{\begin{aligned} 
\lambda_d(1-r^2_w)+2\nu_A \lambda_{AA}\alpha_{AA}(1-\alpha_{AA})
+\lambda_{AB}\nu_B X_B(2-X_B)-X^2_AX^2_B \nu_A \nu_B \lambda^2_{AB}\mathcal{Q}
\end{aligned} }, \label{ap: mean EA one-driven}\\
&E_B =  E_A \lambda_{AB} \nu_A X^2_A\mathcal{Q}, \label{ap: mean EB one-driven}
\end{align}
where
\be
\mathcal{Q}=\frac{1}{(2-X_A) X_A \nu_A \lambda_{AB}+2\alpha_{BB}\lambda_{BB} \nu_B(1-\alpha_{BB})}. \label{eq:Q}
\ee
\end{widetext}}
\vspace{10pt}
\section{Time evolution of $E_{tot}$ and $E_{diff}$: Bi-dispersed Maxwell gas \label{appendix bidispersed}}
In this appendix, we solve for the time evolution of the quantities $E_{tot}$ and $E_{diff}$ for the case of bi-dispersed Maxwell gas. We consider two cases: when only one component of the gas is driven and another case when both the components of the gas are driven as described in Sections~\ref{appendix one component driven} and \ref{appendix both components driven} respectively.

\subsection{One component is driven \label{appendix one component driven}}
We consider a bi-dispersed Maxwell gas where only one component, say $A$, is driven  with a rate, $\lambda_d$ and with a driving strength, $\sigma$.  Using the definitions of $E_{tot}$ and $E_{diff}$ as given in Eqs.~(\ref{E1 define}) and (\ref{E2 define}) respectively, we can write for the time evolution of these quantities following Eq.~(\ref{time-ev}) as 
\begin{align}
\begin{split}
\frac{dE_{tot}}{dt}= \chi_{11} E_{tot} + \chi_{12} E_{diff} +\lambda_d \sigma^2,\\
\frac{dE_{diff}}{dt}= \chi_{21} E_{tot} + \chi_{22} E_{diff} +\lambda_d \sigma^2, \label{appendix time ev E_tot and E_{diff}}
\end{split}
\end{align}
where
{\small
\begin{align}
\begin{split}
\chi_{11}&=-\frac{R_{11}+R_{12}+R_{21}+R_{22}}{2},\\
\chi_{12}&=-\frac{R_{11}-R_{12}+R_{21}-R_{22}}{2},\\
\chi_{21}&=-\frac{R_{11}+R_{12}-R_{21}-R_{22}}{2},\\
\chi_{22}&=-\frac{R_{11}-R_{12}-R_{21}+R_{22}}{2}.
\end{split} \label{appendix : chi}
\end{align}}
Equation~(\ref{appendix time ev E_tot and E_{diff}}) can be represented in a compact matrix form as in Eq.~(\ref{time ev E}). The solution for the time evolution of $E_{tot}$ and $E_{diff}$ are given by
\begin{align}
\begin{split}
E_{tot}(t)&=K_+e^{-\lambda_+t}+ K_-e^{-\lambda_-t}+K_0, \\
E_{diff}(t)&=L_+e^{-\lambda_+t}+ L_-e^{-\lambda_-t}+L_0,
\end{split} \label{app: time ev one component driven}
\end{align}
where the coefficients $K_+, K_-,K_0, L_+, L_-$ and $L_0$ are given by
{\footnotesize
\begin{widetext}
\begin{align}
\begin{split}
K_+&=\frac{1}{\gamma}\Big[ (-\lambda_-+\chi_{11})E_{tot}(0)+\chi_{12}E_{diff}(0)-\frac{\chi_{12}-\lambda_-+\chi_{11})}{\lambda_+}\lambda_d \sigma^2 \Big],~
K_- = \frac{1}{\gamma}\Big[ (\lambda_+-\chi_{11})E_{tot}(0)-\chi_{12} E_{diff}(0)+\frac{\chi_{12}-\lambda_++\chi_{11})}{\lambda_-}\lambda_d \sigma^2 \Big], \\
K_0&=\frac{1}{\gamma}\Big[ \frac{\chi_{12}-(\lambda_--\chi_{11})}{\lambda_+}-\frac{\chi_{12}-(\lambda_+ -\chi_{11})}{\lambda_-}   \Big]\lambda_d \sigma^2,\hspace{50pt} L_0=\frac{1}{\gamma}\Big[ \frac{(\lambda_+-\chi_{11})(\lambda_--\chi_{11})}{\chi_{12}\lambda_+}- \frac{(\lambda_+-\chi_{11})(\lambda_--\chi_{11})}{\chi_{12}\lambda_-}  \Big]\lambda_d \sigma^2, \\
L_+&=\frac{1}{\gamma}\Big[ -\frac{(\lambda_+-\chi_{11})(\lambda_--\chi_{11})}{\chi_{12}}E_{tot}(0) + (\lambda_+ - \chi_{11})E_{diff}(0) -\frac{(\lambda_+-\chi_{11})(\lambda_--\chi_{11})}{\chi_{12}\lambda_+}\lambda_d \sigma^2 \Big], \\
L_-&=\frac{1}{\gamma}\Big[ \frac{(\lambda_+-\chi_{11})(\lambda_--\chi_{11})}{\chi_{12}}E_{tot}(0) - (\lambda_- - \chi_{11})E_{diff}(0) +\frac{(\lambda_+-\chi_{11})(\lambda_--\chi_{11})}{\chi_{12}\lambda_-}\lambda_d \sigma^2 \Big],\\
\gamma&=\lambda_+-\lambda_-. 
\end{split}\label{appendix coeff one component driven}
\end{align}
\end{widetext}}

\subsection{Both components are driven \label{appendix both components driven}}

Here, we consider a bi-dispersed Maxwell gas where both the components, $A$ and $B$ are driven with driving strengths $\sigma_A$ and $\sigma_B$ respectively. One can follow the calculations for the case of one component driven bi-dispersed gas [see subsection \ref{appendix one component driven}] and write the time evolution of $E_{tot}$ and $E_{diff}$ in a compact representation as in Eq.~(\ref{time ev E}) but the column matrix~$\boldsymbol{D}$ takes the form
\begin{align}
\boldsymbol{D}&=\begin{bmatrix}
\lambda_d~(\sigma^2_A+\sigma^2_B), &
\lambda_d~(\sigma^2_A-\sigma^2_B)
\end{bmatrix}^T.
\end{align}
The solution for  $E_{tot}(t)$ and $E_{diff}(t)$ are given by Eq.~(\ref{app: time ev one component driven}) with the
 coefficients $K_+, K_-,K_0, L_+,L_-$ and $L_0$ in Eq.~(\ref{app: time ev one component driven})  given by
 {\small
\begin{widetext}
\begin{align}
K_+&=\frac{1}{\gamma}\Big[ -(\lambda_--\chi_{11})E_{tot}(0)+\chi_{12}E_{diff}(0)-\frac{\lambda_d}{\lambda_+}\big[\big( \chi_{12}-(\lambda_--\chi_{11})  \big)\sigma^2_A - \big( \chi_{12}+(\lambda_--\chi_{11})  \big) \sigma^2_B \big] \Big], \nonumber\\
K_-&= \frac{1}{\gamma}\Big[ (\lambda_+-\chi_{11})E_{tot}(0)-\chi_{12}E_{diff}(0)+ \frac{\lambda_d}{\lambda_-}\big[\big( \chi_{12}-(\lambda_+-\chi_{11})  \big)\sigma^2_A - \big( \chi_{12}+(\lambda_+-\chi_{11})  \big) \sigma^2_B \big] \Big],\nonumber \\
K_0&=\frac{\lambda_d}{\gamma}\Big[ \frac{\big( \chi_{12}-(\lambda_--\chi_{11})  \big)\sigma^2_A - \big( \chi_{12}+(\lambda_--\chi_{11})  \big) \sigma^2_B}{\lambda_+} - \frac{\big( \chi_{12}-(\lambda_+-\chi_{11})  \big)\sigma^2_A - \big( \chi_{12}+(\lambda_+-\chi_{11})  \big) \sigma^2_B}{\lambda_-} \Big],\nonumber\\
L_+&=\frac{1}{\gamma}\Big[ -\frac{(\lambda_+-\chi_{11})(\lambda_--\chi_{11})}{\chi_{12}}E_{tot}(0) + (\lambda_+ - \chi_{11})E_{diff}(0) -\frac{\lambda_d}{\lambda_+ \chi_{12}}\big[(\lambda_+-\chi_{11})(\lambda_--\chi_{11}) (\sigma^2_A- \sigma^2_B) \big] \Big], \nonumber\\ 
L_-&=\frac{1}{\gamma}\Big[ \frac{(\lambda_+-\chi_{11})(\lambda_--\chi_{11})}{\chi_{12}}E_{tot}(0) - (\lambda_+ - \chi_{11})E_{diff}(0) +\frac{\lambda_d}{\lambda_- \chi_{12}}\big[(\lambda_+-\chi_{11})(\lambda_--\chi_{11}) (\sigma^2_A- \sigma^2_B) \big] \Big], \nonumber\\ 
L_0&=\frac{\lambda_d}{\chi_{12}\gamma} \Big[ (\lambda_+-\chi_{11})(\lambda_--\chi_{11}) (\sigma^2_A- \sigma^2_B) \big( \frac{1}{\lambda_+}-\frac{1}{\lambda_-} \big) \Big]            , \nonumber\\ 
\gamma&=\lambda_+-\lambda_-.    \label{appendix coeff both components driven}
\end{align}
\end{widetext}}

\section{Steady state expression for $\Delta E_{tot}/\Delta E_{diff}$ \label{steady state exp}}
In this appendix, we calculate the steady state expression for the ratio $\Delta E_{tot}/\Delta E_{diff}$ for a  bi-dispersed Maxwell gas where both the components, $A$ and $B$ are driven at a rate $\lambda_d$ and with driving strengths $\sigma_A$ and $\sigma_B$ respectively.
The quantities $\Delta E_{tot}$ and $\Delta E_{diff}$ are defined as in Eq.~(\ref{eq: delta E}).
Using the results for the steady state mean kinetic energies for components $A$ and $B$ [see Eqs.~(\ref{66}) and (\ref{67})], one can calculate the ratio $\Delta E_{tot}/\Delta E_{diff}$ as
\begin{widetext}
{\small 
\begin{align}
\frac{\Delta E_{tot}}{\Delta E_{diff}}=\frac{\begin{aligned}
& \lambda_{AB}\lambda_d\Big[\nu_B X^2_B ((\sigma^P_B)^2-(\sigma^Q_B)^2))+\nu_A X^2_A ((\sigma^P_A)^2-(\sigma^Q_A)^2)\Big] \\ &-\lambda_d \Big[-\lambda_d(1-r^2_w)-2 \lambda_{AB} \nu_A X_A + \lambda_{AB} \nu_A X^2_A-\lambda_{BB} \nu_B \alpha_{BB} (1-\alpha_{BB}) \Big]((\sigma^P_A)^2-(\sigma^Q_A)^2) \\ &-\lambda_d \Big[-\lambda_d(1-r^2_w)-2 \lambda_{AB} \nu_B X_B + \lambda_{AB} \nu_B X^2_B-\lambda_{AA} \nu_A \alpha_{AA} (1-\alpha_{AA}) \Big]((\sigma^P_B)^2-(\sigma^Q_B)^2)
\end{aligned}}{\begin{aligned}
& \lambda_{AB}\lambda_d\Big[\nu_B X^2_B ((\sigma^P_B)^2-(\sigma^Q_B)^2))-\nu_A X^2_A ((\sigma^P_A)^2-(\sigma^Q_A)^2)\Big] \\ &-\lambda_d \Big[-\lambda_d(1-r^2_w)-2 \lambda_{AB} \nu_A X_A + \lambda_{AB} \nu_A X^2_A-\lambda_{BB} \nu_B \alpha_{BB} (1-\alpha_{BB}) \Big]((\sigma^P_A)^2-(\sigma^Q_A)^2) \\ &+\lambda_d \Big[-\lambda_d(1-r^2_w)-2 \lambda_{AB} \nu_B X_B + \lambda_{AB} \nu_B X^2_B-\lambda_{AA} \nu_A \alpha_{AA} (1-\alpha_{AA}) \Big]((\sigma^P_B)^2-(\sigma^Q_B)^2)
\end{aligned}}.   \label{critical condition ME both driven}
\end{align}}

 One can also obtain the ratio of $E_{tot}(0)/E_{diff}(0)$ for a system ($P$) only using Eq.~(\ref{critical condition ME both driven}) as

{\small \begin{align}
\frac{ E^P_{tot}}{E^P_{diff}}=\frac{\begin{aligned}
& \lambda_{AB}\lambda_d\Big[\nu_B X^2_B (\sigma^P_B)^2+\nu_A X^2_A (\sigma^P_A)^2\Big] \\ &-\lambda_d \Big[-\lambda_d(1-r^2_w)-2 \lambda_{AB} \nu_A X_A + \lambda_{AB} \nu_A X^2_A-\lambda_{BB} \nu_B \alpha_{BB} (1-\alpha_{BB}) \Big](\sigma^P_A)^2 \\ &-\lambda_d \Big[-\lambda_d(1-r^2_w)-2 \lambda_{AB} \nu_B X_B + \lambda_{AB} \nu_B X^2_B-\lambda_{AA} \nu_A \alpha_{AA} (1-\alpha_{AA}) \Big](\sigma^P_B)^2
\end{aligned}}{\begin{aligned}
& \lambda_{AB}\lambda_d\Big[\nu_B X^2_B (\sigma^P_B)^2-\nu_A X^2_A (\sigma^P_A)^2\Big] \\ &-\lambda_d \Big[-\lambda_d(1-r^2_w)-2 \lambda_{AB} \nu_A X_A + \lambda_{AB} \nu_A X^2_A-\lambda_{BB} \nu_B \alpha_{BB} (1-\alpha_{BB}) \Big](\sigma^P_A)^2 \\ &+\lambda_d \Big[-\lambda_d(1-r^2_w)-2 \lambda_{AB} \nu_B X_B + \lambda_{AB} \nu_B X^2_B-\lambda_{AA} \nu_A \alpha_{AA} (1-\alpha_{AA}) \Big](\sigma^P_B)^2
\end{aligned}}.   \label{appendix: steady state ratio energies}
\end{align}}
\end{widetext}

%\bibliography{ref-mpemba}

\begin{thebibliography}{42}%
\makeatletter
\providecommand \@ifxundefined [1]{%
 \@ifx{#1\undefined}
}%
\providecommand \@ifnum [1]{%
 \ifnum #1\expandafter \@firstoftwo
 \else \expandafter \@secondoftwo
 \fi
}%
\providecommand \@ifx [1]{%
 \ifx #1\expandafter \@firstoftwo
 \else \expandafter \@secondoftwo
 \fi
}%
\providecommand \natexlab [1]{#1}%
\providecommand \enquote  [1]{``#1''}%
\providecommand \bibnamefont  [1]{#1}%
\providecommand \bibfnamefont [1]{#1}%
\providecommand \citenamefont [1]{#1}%
\providecommand \href@noop [0]{\@secondoftwo}%
\providecommand \href [0]{\begingroup \@sanitize@url \@href}%
\providecommand \@href[1]{\@@startlink{#1}\@@href}%
\providecommand \@@href[1]{\endgroup#1\@@endlink}%
\providecommand \@sanitize@url [0]{\catcode `\\12\catcode `\$12\catcode
  `\&12\catcode `\#12\catcode `\^12\catcode `\_12\catcode `\%12\relax}%
\providecommand \@@startlink[1]{}%
\providecommand \@@endlink[0]{}%
\providecommand \url  [0]{\begingroup\@sanitize@url \@url }%
\providecommand \@url [1]{\endgroup\@href {#1}{\urlprefix }}%
\providecommand \urlprefix  [0]{URL }%
\providecommand \Eprint [0]{\href }%
\providecommand \doibase [0]{http://dx.doi.org/}%
\providecommand \selectlanguage [0]{\@gobble}%
\providecommand \bibinfo  [0]{\@secondoftwo}%
\providecommand \bibfield  [0]{\@secondoftwo}%
\providecommand \translation [1]{[#1]}%
\providecommand \BibitemOpen [0]{}%
\providecommand \bibitemStop [0]{}%
\providecommand \bibitemNoStop [0]{.\EOS\space}%
\providecommand \EOS [0]{\spacefactor3000\relax}%
\providecommand \BibitemShut  [1]{\csname bibitem#1\endcsname}%
\let\auto@bib@innerbib\@empty
%</preamble>
\bibitem [{\citenamefont {Aristotle}(0BCE)}]{aristotle}%
  \BibitemOpen
  \bibfield  {author} {\bibinfo {author} {\bibnamefont {Aristotle}},\ }\href
  {http://classics.mit.edu/Aristotle/metaphysics.html} {\emph {\bibinfo {title}
  {Metaphysics}}},\ edited by\ \bibinfo {editor} {\bibfnamefont
  {T.}~\bibnamefont {by~W.~D.~Ross}}\ (\bibinfo  {publisher} {The Internet
  Classics Archive},\ \bibinfo {year} {350BCE})\BibitemShut {NoStop}%
\bibitem [{\citenamefont {Lee}(1952)}]{1952meteorologica}%
  \BibitemOpen
  \bibfield  {author} {\bibinfo {author} {\bibfnamefont {H.}~\bibnamefont
  {Lee}},\ }\href {https://books.google.co.in/books?id=0-iwAAAAIAAJ} {\emph
  {\bibinfo {title} {Meteorologica:}}},\ (Loeb classical library. Greek
  authors.)\ (\bibinfo  {publisher} {Harvard University Press},\ \bibinfo
  {year} {1952})\BibitemShut {NoStop}%
\bibitem [{\citenamefont {Mpemba}\ and\ \citenamefont
  {Osborne}(1969)}]{Mpemba_1969}%
  \BibitemOpen
  \bibfield  {author} {\bibinfo {author} {\bibfnamefont {E.~B.}\ \bibnamefont
  {Mpemba}}\ and\ \bibinfo {author} {\bibfnamefont {D.~G.}\ \bibnamefont
  {Osborne}},\ }\href {\doibase 10.1088/0031-9120/4/3/312} {\bibfield
  {journal} {\bibinfo  {journal} {Physics Education}\ }\textbf {\bibinfo
  {volume} {4}},\ \bibinfo {pages} {172} (\bibinfo {year} {1969})}\BibitemShut
  {NoStop}%
\bibitem [{\citenamefont {Auerbach}(1995)}]{david-super-cooling-1995}%
  \BibitemOpen
  \bibfield  {author} {\bibinfo {author} {\bibfnamefont {D.}~\bibnamefont
  {Auerbach}},\ }\href {\doibase 10.1119/1.18059} {\bibfield  {journal}
  {\bibinfo  {journal} {American Journal of Physics}\ }\textbf {\bibinfo
  {volume} {63}},\ \bibinfo {pages} {882} (\bibinfo {year} {1995})}\BibitemShut
  {NoStop}%
\bibitem [{\citenamefont {Vynnycky}\ and\ \citenamefont
  {Kimura}(2015)}]{vynnycky-convection:2015}%
  \BibitemOpen
  \bibfield  {author} {\bibinfo {author} {\bibfnamefont {M.}~\bibnamefont
  {Vynnycky}}\ and\ \bibinfo {author} {\bibfnamefont {S.}~\bibnamefont
  {Kimura}},\ }\href {\doibase
  https://doi.org/10.1016/j.ijheatmasstransfer.2014.09.015} {\bibfield
  {journal} {\bibinfo  {journal} {International Journal of Heat and Mass
  Transfer}\ }\textbf {\bibinfo {volume} {80}},\ \bibinfo {pages} {243 }
  (\bibinfo {year} {2015})}\BibitemShut {NoStop}%
\bibitem [{\citenamefont {Mirabedin}\ and\ \citenamefont
  {Farhadi}(2017)}]{Mirabedin-evporation-2017}%
  \BibitemOpen
  \bibfield  {author} {\bibinfo {author} {\bibfnamefont {S.~M.}\ \bibnamefont
  {Mirabedin}}\ and\ \bibinfo {author} {\bibfnamefont {F.}~\bibnamefont
  {Farhadi}},\ }\href {\doibase https://doi.org/10.1016/j.ijrefrig.2016.09.006}
  {\bibfield  {journal} {\bibinfo  {journal} {International Journal of
  Refrigeration}\ }\textbf {\bibinfo {volume} {73}},\ \bibinfo {pages} {219 }
  (\bibinfo {year} {2017})}\BibitemShut {NoStop}%
\bibitem [{\citenamefont {Zhang}\ \emph {et~al.}(2014)\citenamefont {Zhang},
  \citenamefont {Huang}, \citenamefont {Ma}, \citenamefont {Zhou},
  \citenamefont {Zhou}, \citenamefont {Zheng}, \citenamefont {Jiang},\ and\
  \citenamefont {Sun}}]{zhang-hydrbond1-2014}%
  \BibitemOpen
  \bibfield  {author} {\bibinfo {author} {\bibfnamefont {X.}~\bibnamefont
  {Zhang}}, \bibinfo {author} {\bibfnamefont {Y.}~\bibnamefont {Huang}},
  \bibinfo {author} {\bibfnamefont {Z.}~\bibnamefont {Ma}}, \bibinfo {author}
  {\bibfnamefont {Y.}~\bibnamefont {Zhou}}, \bibinfo {author} {\bibfnamefont
  {J.}~\bibnamefont {Zhou}}, \bibinfo {author} {\bibfnamefont {W.}~\bibnamefont
  {Zheng}}, \bibinfo {author} {\bibfnamefont {Q.}~\bibnamefont {Jiang}}, \ and\
  \bibinfo {author} {\bibfnamefont {C.~Q.}\ \bibnamefont {Sun}},\ }\href
  {\doibase 10.1039/C4CP03669G} {\bibfield  {journal} {\bibinfo  {journal}
  {Phys. Chem. Chem. Phys.}\ }\textbf {\bibinfo {volume} {16}},\ \bibinfo
  {pages} {22995} (\bibinfo {year} {2014})}\BibitemShut {NoStop}%
\bibitem [{\citenamefont {Tao}\ \emph {et~al.}(2017)\citenamefont {Tao},
  \citenamefont {Zou}, \citenamefont {Jia}, \citenamefont {Li},\ and\
  \citenamefont {Cremer}}]{tao-hydrogen-2017}%
  \BibitemOpen
  \bibfield  {author} {\bibinfo {author} {\bibfnamefont {Y.}~\bibnamefont
  {Tao}}, \bibinfo {author} {\bibfnamefont {W.}~\bibnamefont {Zou}}, \bibinfo
  {author} {\bibfnamefont {J.}~\bibnamefont {Jia}}, \bibinfo {author}
  {\bibfnamefont {W.}~\bibnamefont {Li}}, \ and\ \bibinfo {author}
  {\bibfnamefont {D.}~\bibnamefont {Cremer}},\ }\href {\doibase
  10.1021/acs.jctc.6b00735} {\bibfield  {journal} {\bibinfo  {journal} {Journal
  of Chemical Theory and Computation}\ }\textbf {\bibinfo {volume} {13}},\
  \bibinfo {pages} {55} (\bibinfo {year} {2017})}\BibitemShut {NoStop}%
\bibitem [{\citenamefont {Burridge}\ and\ \citenamefont
  {Linden}(2016)}]{NoMpemba}%
  \BibitemOpen
  \bibfield  {author} {\bibinfo {author} {\bibfnamefont {H.~C.}\ \bibnamefont
  {Burridge}}\ and\ \bibinfo {author} {\bibfnamefont {P.~F.}\ \bibnamefont
  {Linden}},\ }\href@noop {} {\bibfield  {journal} {\bibinfo  {journal}
  {Scientific Reports}\ }\textbf {\bibinfo {volume} {6}},\
  \bibinfo {pages} {37665} (\bibinfo {year}
  {2016})}\BibitemShut {NoStop}%
\bibitem [{\citenamefont {Chaddah}\ \emph {et~al.}(2010)\citenamefont
  {Chaddah}, \citenamefont {Dash}, \citenamefont {Kumar},\ and\ \citenamefont
  {Banerjee}}]{chaddah2010overtaking}%
  \BibitemOpen
  \bibfield  {author} {\bibinfo {author} {\bibfnamefont {P.}~\bibnamefont
  {Chaddah}}, \bibinfo {author} {\bibfnamefont {S.}~\bibnamefont {Dash}},
  \bibinfo {author} {\bibfnamefont {K.}~\bibnamefont {Kumar}}, \ and\ \bibinfo
  {author} {\bibfnamefont {A.}~\bibnamefont {Banerjee}},\ }\href@noop {}
  {\bibfield  {journal} {\bibinfo  {journal} {arXiv preprint arXiv:1011.3598}\
  } (\bibinfo {year} {2010})}\BibitemShut {NoStop}%
\bibitem [{\citenamefont {Ahn}\ \emph {et~al.}(2016)\citenamefont {Ahn},
  \citenamefont {Kang}, \citenamefont {Koh},\ and\ \citenamefont
  {Lee}}]{paper:hydrates}%
  \BibitemOpen
  \bibfield  {author} {\bibinfo {author} {\bibfnamefont {Y.-H.}\ \bibnamefont
  {Ahn}}, \bibinfo {author} {\bibfnamefont {H.}~\bibnamefont {Kang}}, \bibinfo
  {author} {\bibfnamefont {D.-Y.}\ \bibnamefont {Koh}}, \ and\ \bibinfo
  {author} {\bibfnamefont {H.}~\bibnamefont {Lee}},\ }\href@noop {} {\bibfield
  {journal} {\bibinfo  {journal} {Korean Journal of Chemical Engineering}\ \textbf {\bibinfo {volume} {33}} ,\
  \bibinfo {pages} {1903-1907}} (\bibinfo {year} {2016})}\BibitemShut {NoStop}%
\bibitem [{\citenamefont {Hu}\ \emph {et~al.}(2018)\citenamefont {Hu},
  \citenamefont {Li}, \citenamefont {Huang}, \citenamefont {Li}, \citenamefont
  {Luo}, \citenamefont {Chen}, \citenamefont {Jiang},\ and\ \citenamefont
  {An}}]{Polylactide}%
  \BibitemOpen
  \bibfield  {author} {\bibinfo {author} {\bibfnamefont {C.}~\bibnamefont
  {Hu}}, \bibinfo {author} {\bibfnamefont {J.}~\bibnamefont {Li}}, \bibinfo
  {author} {\bibfnamefont {S.}~\bibnamefont {Huang}}, \bibinfo {author}
  {\bibfnamefont {H.}~\bibnamefont {Li}}, \bibinfo {author} {\bibfnamefont
  {C.}~\bibnamefont {Luo}}, \bibinfo {author} {\bibfnamefont {J.}~\bibnamefont
  {Chen}}, \bibinfo {author} {\bibfnamefont {S.}~\bibnamefont {Jiang}}, \ and\
  \bibinfo {author} {\bibfnamefont {L.}~\bibnamefont {An}},\ }\href {\doibase 10.1021/acs.cgd.8b01250
}
  {\bibfield  {journal} {\bibinfo  {journal} {Crystal Growth \& Design}\
  }\textbf {\bibinfo {volume} {18}}(10),\ \bibinfo {pages} {5757} (\bibinfo {year}
  {2018})}\BibitemShut {NoStop}%
\bibitem [{\citenamefont {Gij{\'o}n}\ \emph {et~al.}(2019)\citenamefont
  {Gij{\'o}n}, \citenamefont {Lasanta},\ and\ \citenamefont
  {Hern{\'a}ndez}}]{gijon2019paths}%
  \BibitemOpen
  \bibfield  {author} {\bibinfo {author} {\bibfnamefont {A.}~\bibnamefont
  {Gij{\'o}n}}, \bibinfo {author} {\bibfnamefont {A.}~\bibnamefont {Lasanta}},
  \ and\ \bibinfo {author} {\bibfnamefont {E.}~\bibnamefont {Hern{\'a}ndez}},\
  }\href@noop {} {\bibfield  {journal} {\bibinfo  {journal} {Physical Review
  E}\ }\textbf {\bibinfo {volume} {100}},\ \bibinfo {pages} {032103} (\bibinfo
  {year} {2019})}\BibitemShut {NoStop}%
\bibitem [{\citenamefont {Jin}\ and\ \citenamefont
  {Goddard~III}(2015)}]{Molecular_Dynamics_jin2015mechanisms}%
  \BibitemOpen
  \bibfield  {author} {\bibinfo {author} {\bibfnamefont {J.}~\bibnamefont
  {Jin}}\ and\ \bibinfo {author} {\bibfnamefont {W.~A.}\ \bibnamefont
  {Goddard~III}},\ }\href@noop {} {\bibfield  {journal} {\bibinfo  {journal}
  {The Journal of Physical Chemistry C}\ }\textbf {\bibinfo {volume} {119}},\
  \bibinfo {pages} {2622} (\bibinfo {year} {2015})}\BibitemShut {NoStop}%
\bibitem [{\citenamefont {Gal}\ and\ \citenamefont
  {Raz}(2020)}]{PhysRevLett.124.060602}%
  \BibitemOpen
  \bibfield  {author} {\bibinfo {author} {\bibfnamefont {A.}~\bibnamefont
  {Gal}}\ and\ \bibinfo {author} {\bibfnamefont {O.}~\bibnamefont {Raz}},\
  }\href {\doibase 10.1103/PhysRevLett.124.060602} {\bibfield  {journal}
  {\bibinfo  {journal} {Phys. Rev. Lett.}\ }\textbf {\bibinfo {volume} {124}},\
  \bibinfo {pages} {060602} (\bibinfo {year} {2020})}\BibitemShut {NoStop}%
\bibitem [{\citenamefont {Baity-Jesi}\ \emph {et~al.}(2019)\citenamefont
  {Baity-Jesi}, \citenamefont {Calore}, \citenamefont {Cruz}, \citenamefont
  {Fernandez}, \citenamefont {Gil-Narvi{\'o}n}, \citenamefont
  {Gordillo-Guerrero}, \citenamefont {I{\~n}iguez}, \citenamefont {Lasanta},
  \citenamefont {Maiorano}, \citenamefont {Marinari} \emph
  {et~al.}}]{SpinGlassMpemba}%
  \BibitemOpen
  \bibfield  {author} {\bibinfo {author} {\bibfnamefont {M.}~\bibnamefont
  {Baity-Jesi}}, \bibinfo {author} {\bibfnamefont {E.}~\bibnamefont {Calore}},
  \bibinfo {author} {\bibfnamefont {A.}~\bibnamefont {Cruz}}, \bibinfo {author}
  {\bibfnamefont {L.~A.}\ \bibnamefont {Fernandez}}, \bibinfo {author}
  {\bibfnamefont {J.~M.}\ \bibnamefont {Gil-Narvi{\'o}n}}, \bibinfo {author}
  {\bibfnamefont {A.}~\bibnamefont {Gordillo-Guerrero}}, \bibinfo {author}
  {\bibfnamefont {D.}~\bibnamefont {I{\~n}iguez}}, \bibinfo {author}
  {\bibfnamefont {A.}~\bibnamefont {Lasanta}}, \bibinfo {author} {\bibfnamefont
  {A.}~\bibnamefont {Maiorano}}, \bibinfo {author} {\bibfnamefont
  {E.}~\bibnamefont {Marinari}},  \emph {et~al.},\ }\href@noop {} {\bibfield
  {journal} {\bibinfo  {journal} {Proceedings of the National Academy of
  Sciences}\ }\textbf {\bibinfo {volume} {116}},\ \bibinfo {pages} {15350}
  (\bibinfo {year} {2019})}\BibitemShut {NoStop}%
\bibitem [{\citenamefont {Lu}\ and\ \citenamefont {Raz}(2017)}]{Lu-raz:2017}%
  \BibitemOpen
  \bibfield  {author} {\bibinfo {author} {\bibfnamefont {Z.}~\bibnamefont
  {Lu}}\ and\ \bibinfo {author} {\bibfnamefont {O.}~\bibnamefont {Raz}},\
  }\href {\doibase 10.1073/pnas.1701264114} {\bibfield  {journal} {\bibinfo
  {journal} {Proceedings of the National Academy of Sciences}\ }\textbf
  {\bibinfo {volume} {114}},\ \bibinfo {pages} {5083} (\bibinfo {year}
  {2017})}\BibitemShut {NoStop}%
\bibitem [{\citenamefont {Klich}\ \emph {et~al.}(2019)\citenamefont {Klich},
  \citenamefont {Raz}, \citenamefont {Hirschberg},\ and\ \citenamefont
  {Vucelja}}]{Klich-2019}%
  \BibitemOpen
  \bibfield  {author} {\bibinfo {author} {\bibfnamefont {I.}~\bibnamefont
  {Klich}}, \bibinfo {author} {\bibfnamefont {O.}~\bibnamefont {Raz}}, \bibinfo
  {author} {\bibfnamefont {O.}~\bibnamefont {Hirschberg}}, \ and\ \bibinfo
  {author} {\bibfnamefont {M.}~\bibnamefont {Vucelja}},\ }\href {\doibase
  10.1103/PhysRevX.9.021060} {\bibfield  {journal} {\bibinfo  {journal} {Phys.
  Rev. X}\ }\textbf {\bibinfo {volume} {9}},\ \bibinfo {pages} {021060}
  (\bibinfo {year} {2019})}\BibitemShut {NoStop}%
\bibitem [{\citenamefont {Klich}\ and\ \citenamefont
  {Vucelja}(2018)}]{klich2018solution}%
  \BibitemOpen
  \bibfield  {author} {\bibinfo {author} {\bibfnamefont {I.}~\bibnamefont
  {Klich}}\ and\ \bibinfo {author} {\bibfnamefont {M.}~\bibnamefont
  {Vucelja}},\ }\href@noop {} {\bibfield  {journal} {\bibinfo  {journal} {arXiv
  preprint arXiv:1812.11962}\ } (\bibinfo {year} {2018})}\BibitemShut {NoStop}%
\bibitem [{\citenamefont {Lasanta}\ \emph {et~al.}(2017)\citenamefont
  {Lasanta}, \citenamefont {Vega~Reyes}, \citenamefont {Prados},\ and\
  \citenamefont {Santos}}]{Lasanta-mpemba-1-2017}%
  \BibitemOpen
  \bibfield  {author} {\bibinfo {author} {\bibfnamefont {A.}~\bibnamefont
  {Lasanta}}, \bibinfo {author} {\bibfnamefont {F.}~\bibnamefont {Vega~Reyes}},
  \bibinfo {author} {\bibfnamefont {A.}~\bibnamefont {Prados}}, \ and\ \bibinfo
  {author} {\bibfnamefont {A.}~\bibnamefont {Santos}},\ }\href {\doibase
  10.1103/PhysRevLett.119.148001} {\bibfield  {journal} {\bibinfo  {journal}
  {Phys. Rev. Lett.}\ }\textbf {\bibinfo {volume} {119}},\ \bibinfo {pages}
  {148001} (\bibinfo {year} {2017})}\BibitemShut {NoStop}%
\bibitem [{\citenamefont {Torrente}\ \emph {et~al.}(2019)\citenamefont
  {Torrente}, \citenamefont {L\'opez-Casta\~no}, \citenamefont {Lasanta},
  \citenamefont {Reyes}, \citenamefont {Prados},\ and\ \citenamefont
  {Santos}}]{Torrente-rough-2019}%
  \BibitemOpen
  \bibfield  {author} {\bibinfo {author} {\bibfnamefont {A.}~\bibnamefont
  {Torrente}}, \bibinfo {author} {\bibfnamefont {M.~A.}\ \bibnamefont
  {L\'opez-Casta\~no}}, \bibinfo {author} {\bibfnamefont {A.}~\bibnamefont
  {Lasanta}}, \bibinfo {author} {\bibfnamefont {F.~V.}\ \bibnamefont {Reyes}},
  \bibinfo {author} {\bibfnamefont {A.}~\bibnamefont {Prados}}, \ and\ \bibinfo
  {author} {\bibfnamefont {A.}~\bibnamefont {Santos}},\ }\href {\doibase
  10.1103/PhysRevE.99.060901} {\bibfield  {journal} {\bibinfo  {journal} {Phys.
  Rev. E}\ }\textbf {\bibinfo {volume} {99}},\ \bibinfo {pages} {060901}
  (\bibinfo {year} {2019})}\BibitemShut {NoStop}%
\bibitem [{\citenamefont {Momp{\'o}}\ \emph {et~al.}(2020)\citenamefont
  {Momp{\'o}}, \citenamefont {Casta{\~n}o}, \citenamefont {Torrente},
  \citenamefont {Reyes},\ and\ \citenamefont {Lasanta}}]{mompo2020memory}%
  \BibitemOpen
  \bibfield  {author} {\bibinfo {author} {\bibfnamefont {E.}~\bibnamefont
  {Momp{\'o}}}, \bibinfo {author} {\bibfnamefont {M.}~\bibnamefont
  {Casta{\~n}o}}, \bibinfo {author} {\bibfnamefont {A.}~\bibnamefont
  {Torrente}}, \bibinfo {author} {\bibfnamefont {F.~V.}\ \bibnamefont {Reyes}},
  \ and\ \bibinfo {author} {\bibfnamefont {A.}~\bibnamefont {Lasanta}},\
  }\href@noop {} {\bibfield  {journal} {\bibinfo  {journal} {arXiv preprint
  arXiv:2006.00241}\ } (\bibinfo {year} {2020})}\BibitemShut {NoStop}%
\bibitem [{\citenamefont {Ben-Naim}\ and\ \citenamefont
  {Krapivsky}(2000)}]{Ben-naim:00}%
  \BibitemOpen
  \bibfield  {author} {\bibinfo {author} {\bibfnamefont {E.}~\bibnamefont
  {Ben-Naim}}\ and\ \bibinfo {author} {\bibfnamefont {P.~L.}\ \bibnamefont
  {Krapivsky}},\ }\href {\doibase 10.1103/PhysRevE.61.R5} {\bibfield  {journal}
  {\bibinfo  {journal} {Phys. Rev. E}\ }\textbf {\bibinfo {volume} {61}},\
  \bibinfo {pages} {R5} (\bibinfo {year} {2000})}\BibitemShut {NoStop}%
\bibitem [{\citenamefont {Bobylev}\ \emph {et~al.}(2000)\citenamefont
  {Bobylev}, \citenamefont {Carrillo},\ and\ \citenamefont
  {Gamba}}]{Bobylev:00}%
  \BibitemOpen
  \bibfield  {author} {\bibinfo {author} {\bibfnamefont {A.~V.}\ \bibnamefont
  {Bobylev}}, \bibinfo {author} {\bibfnamefont {J.~A.}\ \bibnamefont
  {Carrillo}}, \ and\ \bibinfo {author} {\bibfnamefont {I.~M.}\ \bibnamefont
  {Gamba}},\ }\href {\doibase 10.1023/A:1018627625800} {\bibfield  {journal}
  {\bibinfo  {journal} {J. Stat. Phys.}\ }\textbf {\bibinfo {volume} {98}},\
  \bibinfo {pages} {743} (\bibinfo {year} {2000})}\BibitemShut {NoStop}%
\bibitem [{\citenamefont {Prasad}\ \emph
  {et~al.}(2014{\natexlab{a}})\citenamefont {Prasad}, \citenamefont
  {Sabhapandit},\ and\ \citenamefont {Dhar}}]{prasad2014high}%
  \BibitemOpen
  \bibfield  {author} {\bibinfo {author} {\bibfnamefont {V.~V.}~\bibnamefont
  {Prasad}}, \bibinfo {author} {\bibfnamefont {S.}~\bibnamefont {Sabhapandit}},
  \ and\ \bibinfo {author} {\bibfnamefont {A.}~\bibnamefont {Dhar}},\ }\href
  {\doibase 10.1209/0295-5075/104/54003} {\bibfield  {journal} {\bibinfo
  {journal} {EPL (Europhysics Letters)}\ }\textbf {\bibinfo {volume} {104}},\
  \bibinfo {pages} {54003} (\bibinfo {year} {2014}{\natexlab{a}})}\BibitemShut
  {NoStop}%
\bibitem [{\citenamefont {Prasad}\ \emph
  {et~al.}(2014{\natexlab{b}})\citenamefont {Prasad}, \citenamefont
  {Sabhapandit},\ and\ \citenamefont {Dhar}}]{Prasad:14}%
  \BibitemOpen
  \bibfield  {author} {\bibinfo {author} {\bibfnamefont {V.~V.}\ \bibnamefont
  {Prasad}}, \bibinfo {author} {\bibfnamefont {S.}~\bibnamefont {Sabhapandit}},
  \ and\ \bibinfo {author} {\bibfnamefont {A.}~\bibnamefont {Dhar}},\ }\href
  {\doibase 10.1103/PhysRevE.90.062130} {\bibfield  {journal} {\bibinfo
  {journal} {Phys. Rev. E}\ }\textbf {\bibinfo {volume} {90}},\ \bibinfo
  {pages} {062130} (\bibinfo {year} {2014}{\natexlab{b}})}\BibitemShut
  {NoStop}%
\bibitem [{\citenamefont {Biswas}\ \emph {et~al.}(2020)\citenamefont {Biswas},
  \citenamefont {Prasad},\ and\ \citenamefont {Rajesh}}]{Biswas_2020}%
  \BibitemOpen
  \bibfield  {author} {\bibinfo {author} {\bibfnamefont {A.}~\bibnamefont
  {Biswas}}, \bibinfo {author} {\bibfnamefont {V.~V.}\ \bibnamefont {Prasad}},
  \ and\ \bibinfo {author} {\bibfnamefont {R.}~\bibnamefont {Rajesh}},\ }\href
  {\doibase 10.1088/1742-5468/ab6095} {\bibfield  {journal} {\bibinfo
  {journal} {Journal of Statistical Mechanics: Theory and Experiment}\ }\textbf
  {\bibinfo {volume} {2020}},\ \bibinfo {pages} {013202} (\bibinfo {year}
  {2020})}\BibitemShut {NoStop}%
\bibitem [{\citenamefont {Prasad}\ \emph {et~al.}(2017)\citenamefont {Prasad},
  \citenamefont {Das}, \citenamefont {Sabhapandit},\ and\ \citenamefont
  {Rajesh}}]{prasad2017velocity}%
  \BibitemOpen
  \bibfield  {author} {\bibinfo {author} {\bibfnamefont {V.~V.}~\bibnamefont
  {Prasad}}, \bibinfo {author} {\bibfnamefont {D.}~\bibnamefont {Das}},
  \bibinfo {author} {\bibfnamefont {S.}~\bibnamefont {Sabhapandit}}, \ and\
  \bibinfo {author} {\bibfnamefont {R.}~\bibnamefont {Rajesh}},\ }\href@noop {}
  {\bibfield  {journal} {\bibinfo  {journal} {Physical Review E}\ }\textbf
  {\bibinfo {volume} {95}},\ \bibinfo {pages} {032909} (\bibinfo {year}
  {2017})}\BibitemShut {NoStop}%
\bibitem [{\citenamefont {Prasad}\ and\ \citenamefont
  {Rajesh}(2019)}]{prasad2019asymptotic}%
  \BibitemOpen
  \bibfield  {author} {\bibinfo {author} {\bibfnamefont {V.~V.}~\bibnamefont
  {Prasad}}\ and\ \bibinfo {author} {\bibfnamefont {R.}~\bibnamefont
  {Rajesh}},\ }\href {\doibase 10.1007/s10955-019-02347-8} {\bibfield
  {journal} {\bibinfo  {journal} {Journal of Statistical Physics}\ }\textbf
  {\bibinfo {volume} {176}},\ \bibinfo {pages} {1409} (\bibinfo {year}
  {2019})}\BibitemShut {NoStop}%
\bibitem [{\citenamefont {Prasad}\ \emph {et~al.}(2019)\citenamefont {Prasad},
  \citenamefont {Das}, \citenamefont {Sabhapandit},\ and\ \citenamefont
  {Rajesh}}]{Prasad_2019}%
  \BibitemOpen
  \bibfield  {author} {\bibinfo {author} {\bibfnamefont {V.~V.}\ \bibnamefont
  {Prasad}}, \bibinfo {author} {\bibfnamefont {D.}~\bibnamefont {Das}},
  \bibinfo {author} {\bibfnamefont {S.}~\bibnamefont {Sabhapandit}}, \ and\
  \bibinfo {author} {\bibfnamefont {R.}~\bibnamefont {Rajesh}},\ }\href
  {\doibase 10.1088/1742-5468/ab11da} {\bibfield  {journal} {\bibinfo
  {journal} {Journal of Statistical Mechanics: Theory and Experiment}\ }\textbf
  {\bibinfo {volume} {2019}},\ \bibinfo {pages} {063201} (\bibinfo {year}
  {2019})}\BibitemShut {NoStop}%
\bibitem [{\citenamefont {Baxter}\ and\ \citenamefont
  {Olafsen}(2003)}]{Baxter:03}%
  \BibitemOpen
  \bibfield  {author} {\bibinfo {author} {\bibfnamefont {G.}~\bibnamefont
  {Baxter}}\ and\ \bibinfo {author} {\bibfnamefont {J.}~\bibnamefont
  {Olafsen}},\ }\href@noop {} {\bibfield  {journal} {\bibinfo  {journal}
  {Nature}\ }\textbf {\bibinfo {volume} {425}},\ \bibinfo {pages} {680}
  (\bibinfo {year} {2003})}\BibitemShut {NoStop}%
\bibitem [{\citenamefont {Baxter}\ and\ \citenamefont
  {Olafsen}(2007{\natexlab{a}})}]{baxter2007temperature}%
  \BibitemOpen
  \bibfield  {author} {\bibinfo {author} {\bibfnamefont {G.}~\bibnamefont
  {Baxter}}\ and\ \bibinfo {author} {\bibfnamefont {J.}~\bibnamefont
  {Olafsen}},\ }\href@noop {} {\bibfield  {journal} {\bibinfo  {journal}
  {Granul. Matter}\ }\textbf {\bibinfo {volume} {9}},\ \bibinfo {pages} {135}
  (\bibinfo {year} {2007}{\natexlab{a}})}\BibitemShut {NoStop}%
\bibitem [{\citenamefont {Baxter}\ and\ \citenamefont
  {Olafsen}(2007{\natexlab{b}})}]{Baxter-PRL:2007}%
  \BibitemOpen
  \bibfield  {author} {\bibinfo {author} {\bibfnamefont {G.~W.}\ \bibnamefont
  {Baxter}}\ and\ \bibinfo {author} {\bibfnamefont {J.~S.}\ \bibnamefont
  {Olafsen}},\ }\href {\doibase 10.1103/PhysRevLett.99.028001} {\bibfield
  {journal} {\bibinfo  {journal} {Phys. Rev. Lett.}\ }\textbf {\bibinfo
  {volume} {99}},\ \bibinfo {pages} {028001} (\bibinfo {year}
  {2007}{\natexlab{b}})}\BibitemShut {NoStop}%
\bibitem [{\citenamefont {Combs}\ \emph {et~al.}(2008)\citenamefont {Combs},
  \citenamefont {Olafsen}, \citenamefont {Burdeau},\ and\ \citenamefont
  {Viot}}]{Comb-PRE:2008}%
  \BibitemOpen
  \bibfield  {author} {\bibinfo {author} {\bibfnamefont {K.}~\bibnamefont
  {Combs}}, \bibinfo {author} {\bibfnamefont {J.~S.}\ \bibnamefont {Olafsen}},
  \bibinfo {author} {\bibfnamefont {A.}~\bibnamefont {Burdeau}}, \ and\
  \bibinfo {author} {\bibfnamefont {P.}~\bibnamefont {Viot}},\ }\href {\doibase
  10.1103/PhysRevE.78.042301} {\bibfield  {journal} {\bibinfo  {journal} {Phys.
  Rev. E}\ }\textbf {\bibinfo {volume} {78}},\ \bibinfo {pages} {042301}
  (\bibinfo {year} {2008})}\BibitemShut {NoStop}%
\bibitem [{\citenamefont {Burdeau}\ and\ \citenamefont
  {Viot}(2009)}]{Burdeau-PRE:2009}%
  \BibitemOpen
  \bibfield  {author} {\bibinfo {author} {\bibfnamefont {A.}~\bibnamefont
  {Burdeau}}\ and\ \bibinfo {author} {\bibfnamefont {P.}~\bibnamefont {Viot}},\
  }\href {\doibase 10.1103/PhysRevE.79.061306} {\bibfield  {journal} {\bibinfo
  {journal} {Phys. Rev. E}\ }\textbf {\bibinfo {volume} {79}},\ \bibinfo
  {pages} {061306} (\bibinfo {year} {2009})}\BibitemShut {NoStop}%
\bibitem [{\citenamefont {Windows-Yule}\ and\ \citenamefont
  {Parker}(2013)}]{Yule:2013}%
  \BibitemOpen
  \bibfield  {author} {\bibinfo {author} {\bibfnamefont {C.~R.~K.}\
  \bibnamefont {Windows-Yule}}\ and\ \bibinfo {author} {\bibfnamefont {D.~J.}\
  \bibnamefont {Parker}},\ }\href {\doibase 10.1103/PhysRevE.87.022211}
  {\bibfield  {journal} {\bibinfo  {journal} {Phys. Rev. E}\ }\textbf {\bibinfo
  {volume} {87}},\ \bibinfo {pages} {022211} (\bibinfo {year}
  {2013})}\BibitemShut {NoStop}%
\bibitem [{\citenamefont {Feitosa}\ and\ \citenamefont
  {Menon}(2002)}]{Feitosa2002}%
  \BibitemOpen
  \bibfield  {author} {\bibinfo {author} {\bibfnamefont {K.}~\bibnamefont
  {Feitosa}}\ and\ \bibinfo {author} {\bibfnamefont {N.}~\bibnamefont
  {Menon}},\ }\href {\doibase 10.1103/PhysRevLett.88.198301} {\bibfield
  {journal} {\bibinfo  {journal} {Phys. Rev. Lett.}\ }\textbf {\bibinfo
  {volume} {88}},\ \bibinfo {pages} {198301} (\bibinfo {year}
  {2002})}\BibitemShut {NoStop}%
\bibitem [{\citenamefont {Barrat}\ and\ \citenamefont
  {Trizac}(2002)}]{barrat2002lack}%
  \BibitemOpen
  \bibfield  {author} {\bibinfo {author} {\bibfnamefont {A.}~\bibnamefont
  {Barrat}}\ and\ \bibinfo {author} {\bibfnamefont {E.}~\bibnamefont
  {Trizac}},\ }\href@noop {} {\bibfield  {journal} {\bibinfo  {journal}
  {Granul. Matter}\ }\textbf {\bibinfo {volume} {4}},\ \bibinfo {pages} {57}
  (\bibinfo {year} {2002})}\BibitemShut {NoStop}%
\bibitem [{\citenamefont {Pagnani}\ \emph {et~al.}(2002)\citenamefont
  {Pagnani}, \citenamefont {Bettolo~Marconi},\ and\ \citenamefont
  {Puglisi}}]{Pagnani-PRE:2002}%
  \BibitemOpen
  \bibfield  {author} {\bibinfo {author} {\bibfnamefont {R.}~\bibnamefont
  {Pagnani}}, \bibinfo {author} {\bibfnamefont {U.~M.}\ \bibnamefont
  {Bettolo~Marconi}}, \ and\ \bibinfo {author} {\bibfnamefont {A.}~\bibnamefont
  {Puglisi}},\ }\href {\doibase 10.1103/PhysRevE.66.051304} {\bibfield
  {journal} {\bibinfo  {journal} {Phys. Rev. E}\ }\textbf {\bibinfo {volume}
  {66}},\ \bibinfo {pages} {051304} (\bibinfo {year} {2002})}\BibitemShut
  {NoStop}%
\bibitem [{\citenamefont {Wang}\ and\ \citenamefont {Menon}(2008)}]{Wang2008}%
  \BibitemOpen
  \bibfield  {author} {\bibinfo {author} {\bibfnamefont {H.-Q.}\ \bibnamefont
  {Wang}}\ and\ \bibinfo {author} {\bibfnamefont {N.}~\bibnamefont {Menon}},\
  }\href {\doibase 10.1103/PhysRevLett.100.158001} {\bibfield  {journal}
  {\bibinfo  {journal} {Phys. Rev. Lett.}\ }\textbf {\bibinfo {volume} {100}},\
  \bibinfo {pages} {158001} (\bibinfo {year} {2008})}\BibitemShut {NoStop}%
\bibitem [{\citenamefont {Brey}\ and\ \citenamefont
  {Ruiz-Montero}(2009)}]{JJBrey-non-eq:2009}%
  \BibitemOpen
  \bibfield  {author} {\bibinfo {author} {\bibfnamefont {J.~J.}\ \bibnamefont
  {Brey}}\ and\ \bibinfo {author} {\bibfnamefont {M.~J.}\ \bibnamefont
  {Ruiz-Montero}},\ }\href {\doibase 10.1103/PhysRevE.80.041306} {\bibfield
  {journal} {\bibinfo  {journal} {Phys. Rev. E}\ }\textbf {\bibinfo {volume}
  {80}},\ \bibinfo {pages} {041306} (\bibinfo {year} {2009})}\BibitemShut
  {NoStop}%
\bibitem [{\citenamefont {Uecker}\ \emph {et~al.}(2009)\citenamefont {Uecker},
  \citenamefont {Kranz}, \citenamefont {Aspelmeier},\ and\ \citenamefont
  {Zippelius}}]{Uecker:2009}%
  \BibitemOpen
  \bibfield  {author} {\bibinfo {author} {\bibfnamefont {H.}~\bibnamefont
  {Uecker}}, \bibinfo {author} {\bibfnamefont {W.~T.}\ \bibnamefont {Kranz}},
  \bibinfo {author} {\bibfnamefont {T.}~\bibnamefont {Aspelmeier}}, \ and\
  \bibinfo {author} {\bibfnamefont {A.}~\bibnamefont {Zippelius}},\ }\href
  {\doibase 10.1103/PhysRevE.80.041303} {\bibfield  {journal} {\bibinfo
  {journal} {Phys. Rev. E}\ }\textbf {\bibinfo {volume} {80}},\ \bibinfo
  {pages} {041303} (\bibinfo {year} {2009})}\BibitemShut {NoStop}%
\end{thebibliography}

%

\end{document}